\documentclass[11pt,a4paper]{article} 
\usepackage[dvips]{graphics}
\newcommand{\bold}[1]{\mbox{\boldmath $#1$}} 
\newcommand{\sbold}[1]{\mbox{\boldmath $\scriptstyle #1$}}  
\newcommand{\boldsf}[1]{\mbox{\boldmath $\mathsf #1$}} 
\newcommand{\threej}[6]{\left( \begin{array}{ccc} #1 & #2 & #3 \\ #4 & #5 & #6
                        \end{array} \right) }  
\newcommand{\sixj}[6]{ \left\{ \begin{array}{ccc} #1 & #2 & #3 \\ #4 & #5 & #6
                       \end{array} \right\} }  
\newcommand{\ninej}[9]{ \left\{ \begin{array}{ccc} #1 & #2 & #3 \\ 
#4 & #5 & #6 \\ #7 & #8 & #9 \end{array} \right\} }  
\newcommand{\braket}[2]{\langle\,#1\,|\,#2\,\rangle}  

\newcommand{\Ref}[1]{\mbox{(~\hspace{-.3em}\ref{#1}\hspace{-.3em}~)}}
\newcommand{\mywarn}{******MY WARNING****** : }
\newcommand{\cref}{$\clubsuit$ \marginpar{$\clubsuit${\em Check ref.!}}
\typeout{\mywarn Check reference(s)!}}

\newcommand{\ket}[1]{| #1 \rangle}
\newcommand{\roundket}[1]{\left| #1 \right)}
\newcommand{\bra}[1]{\langle #1 |}
\newcommand{\roundbra}[1]{\left( #1 \right|}
\newcommand{\roundbraket}[2]{\left( #1 \right|\left. #2 \right)} 
\newcommand{\Ket}[1]{|| #1 \rangle}
\newcommand{\Bra}[1]{\langle #1 ||}

\newcommand{\etal}{\emph{et al.}}

\newcommand{\ie}{\emph{i.e.}}
\newcommand{\eg}{\emph{e.g.}}
\newcommand{\appendixeqn}{\setcounter{equation}{0}%
\renewcommand{\theequation}{\mbox{\Alph{section}\arabic{equation}}}}
\newcommand{\spurion}{{0 \choose 1}}
\newcommand{\lhat}[1]{\!\!\!\!\!\hat{\mbox{$\,\,\,\,\,#1$}}}

\newcommand{\none}{\multicolumn{1}{c}{$0$}}
\newcommand{\noneb}{\multicolumn{1}{c|}{$0$}}

\newcommand{\cpos}[1]{\multicolumn{1}{c}{#1}}
\newcommand{\cposb}[1]{\multicolumn{1}{c|}{#1}}
\newlength{\thisheight}
\newlength{\thiswidth}

\begin{document} 
\title{ \bfseries   
Kinematical and nonlocality effects on the nonmesonic weak 
hypernuclear decay 
}
\author{ 
C. Barbero$^1$, 
C. De Conti%
\footnote{Present address: Instituto de Ci\^encias, Universidade Federal de  
Itajub\'a, Avenida BPS 1303/Pi\-nhei\-ri\-nho, 37500-903 Itajub\'a-MG, Brazil.} 
$^{\,2}$, 
A. P. Gale\~ao%
\footnote{Corresponding author: galeao@ift.unesp.br}
$^{\,3}$
and 
F. Krmpoti\'c$^{4,1}$
\\[1ex]
\itshape \normalsize
$^1$ Departamento de F\'{\i}sica, Universidad Nacional de
 La Plata \\
\itshape \normalsize
C.C. 67, 1900 La Plata, Argentina \\
\itshape \normalsize
$^2$ Instituto Tecnol\'ogico de Aeron\'autica, Centro T\'ecnico Aeroespacial \\
\itshape \normalsize
12228-900 S\~ao Jos\'e dos Campos-SP, Brazil \\
\itshape \normalsize
$^3$ Instituto de F\'{\i}sica Te\'orica, Universidade Estadual Paulista \\
\itshape \normalsize
Rua Pamplona 145, 01405-900 S\~ao Paulo-SP, Brazil \\ 
\itshape \normalsize
$^4$ Instituto de F\'{\i}sica, Universidade de S\~ao Paulo \\
\itshape \normalsize
C. P. 66318, 05315-970 S\~ao Paulo-SP, Brazil.
}
\date{23 June 2003}
\maketitle
\vspace{-1\baselineskip}
\begin{abstract}
\noindent
We make a careful study about the nonrelativistic reduction 
of
 one-meson-exchange models for the nonmesonic weak hypernuclear decay.
Starting from a widely accepted effective coupling Hamiltonian involving the 
exchange of the complete pseudoscalar and vector meson octets 
($\pi$, $\eta$, 
$K$, $\rho$, $\omega$, $K^*$), the strangeness-changing weak $\Lambda N \to NN$ 
transition potential
 is derived, 
including  two effects that have been 
systematically omitted in the literature, or, at best, only partly 
considered. These are the
 kinematical effects due to the difference between the 
lambda and
 nucleon masses, and the first-order nonlocality corrections, \ie, 
those involving up to first-order differential operators.  Our analysis clearly 
shows that the main kinematical effect on the local contributions is the 
reduction of the effective pion mass. The kinematical effect 
on the nonlocal contributions is more complicated, since it activates several 
new terms that would otherwise remain dormant. 
Numerical results for 
$^{12}_{\,\;\Lambda}$C and $^5_\Lambda\mathrm{He}$ are presented and they 
show that the combined kinematical 
plus nonlocal corrections have an appreciable influence on  the partial decay 
rates. However, this is somewhat diminished in the main decay observables: the 
total nonmesonic rate, $\Gamma_{nm}$, the neutron-to-proton branching ratio, 
$\Gamma_n/\Gamma_p$, and the asymmetry parameter, $a_\Lambda$. 
The latter two still cannot be reconciled with the available experimental data. 
The existing theoretical predictions for the sign of $a_\Lambda$ in 
$^5_\Lambda\mathrm{He}$ are confirmed.
\end{abstract}
%

%
%
%
\vspace{.5\baselineskip}
\section{Introduction} \label{int}

The free decay of a $\Lambda$ hyperon occurs almost exclusively
through the mesonic mode, $\Lambda \rightarrow \pi N$, with the
nucleon emerging with a momentum of about 100 MeV/c. Inside nuclear
matter ($p_F \approx$ 270 MeV/c)  this mode is Pauli blocked, and, for
all but the lightest $\Lambda$ hypernuclei ($A \ge 5$), the weak
decay is dominated by the nonmesonic channel, $\Lambda N
\rightarrow N N$, which liberates enough kinetic energy to put the two emitted 
nucleons above the Fermi surface. In the absence of stable hyperon beams, these  
nonmesonic decays offer the only way available to investigate the 
strangeness-changing weak interaction between hadrons. 
(For reviews on hypernuclear decay, see Refs.~\cite{Co90}--\cite{Al02}.)

The simplest model for this process is the
exchange of a virtual pion \cite{Ad67}, and in fact this can reproduce
reasonably well the total (nonmesonic) decay rate, $\Gamma_{nm} = \Gamma_n +
\Gamma_p$, but fails badly for other observables like the ratio of
neutron-induced ($\Lambda n \rightarrow n n$) to proton-induced ($\Lambda p 
\rightarrow n p$) transitions, $\Gamma_n/\Gamma_p$, and the asymmetry parameter 
$a_\Lambda$.  
The deficiency of this model is attributed to effects of short range physics, 
which should be quite important in view of the large momentum transfers
involved ($\sim$ 400 MeV/c). Although there have been some attempts to account 
for this fact by making use of quark models to compute the shortest range part 
of the transition potential \cite{Ch83}--\cite{Sa00}, most of the theoretical 
work opted for the addition of other, heavier mesons in the exchange process 
\cite{Mc84}--\cite{Kr02}.  
None of these models gives fully satisfactory results. 
Inclusion of correlated two-pion exchange has not been completely successful 
either \cite{Sh94,It02}. 
Nor have the addition of uncorrelated two-pion exchange, two-nucleon induced 
transitions or medium effects, treated within the nonrelativistic 
\cite{Os85}--\cite{Ji01} or relativistic \cite{Zh99} propagator approaches, been 
of much help.  

Here, we concentrate on the line of one-meson-exchange (OME) models 
\cite{Ad67}--\cite{It02}. 
We do not explicitly discuss hybrid models, \ie, those involving quark degrees 
of freedom \cite{Ch83}--\cite{Sa00}.  However, much of the theoretical 
developments we present could be generalized to include them. Also two-pion 
exchange \cite{Sh94,It02} could be brought into our general framework. 
The main ingredients of OME models are the effective baryon-baryon-meson weak 
and strong Hamiltonians. 
These are constructed in the language of relativistic field theory, but in
almost all calculations (exceptions are Refs.~\cite{Ra91}--\cite{Pa95b}) one
has proceeded to make a nonrelativistic reduction for the extraction of the 
transition potential. 
This often involves some further approximations, 
like neglecting the nonlocality in this
potential, and balancing by hand the distorted kinematics in the
OME Feynman amplitudes, resulting from the difference between
initial and final baryon masses. 
This not only alters the different terms in the  transition potential, but also 
eliminates several of them. 
Our main purpose here is to assess the relative importance of these effects.

The paper is organized as follows. 
Most of the formalism is developed in Section~\ref{OME}. 
In Subsection~\ref{general}, we explain the construction of the nonrelativistic  
transition potential, taking due care of the kinematics and including nonlocal 
terms, and, in Subsection~\ref{Kin}, we give a motivation for not neglecting 
\emph{a~priori} the lambda-nucleon mass difference. 
In Subsections \ref{nstr} and \ref{str}, the explicit 
expressions for the local and first-order nonlocal contributions to the 
transition potential due to the exchange of a $\pi$, $\rho$, $K$ or $K^*$ meson
are derived and commented.  
The ones  corresponding to the $\eta$ and $\omega$ exchanges can be easily 
obtained by analogy with those of the $\pi$ and $\rho$, respectively, thus 
allowing the inclusion of the full pseudoscalar and vector meson octets, as 
deemed necessary by the present day consensus. 
In Subsection~\ref{fse}, we describe how to take the finite size effects into 
account by means of form factors. 
Our numerical results are reported in detail and discussed in Section~\ref{num}. 
The phenomenological way to include short range correlations is presented in 
Subsection~\ref{rates}, together with the main expressions for the calculation 
of the transition rates in the extreme particle-hole model of Ref.~\cite{Ba02}, 
and all this is applied to the decay of $^{12}_{\,\;\Lambda}$C. In 
Subsection~\ref{asymmetry}, we do the same for $^5_\Lambda\mathrm{He}$ and 
compute also the asymmetry parameter.  
Finally,  Section~\ref{con} summarizes our main conclusions. 
Some useful formulas are collected in Appendix~\ref{nuclear} and a specific 
point relating to our phase conventions is discussed in 
Appendix~\ref{conventions}. 
\section{OME transition potential} \label{OME}

\subsection{General discussion} \label{general}

\begin{figure}[htb]
%
\begin{center}
\includegraphics*[5pt,480pt][365pt,790pt]{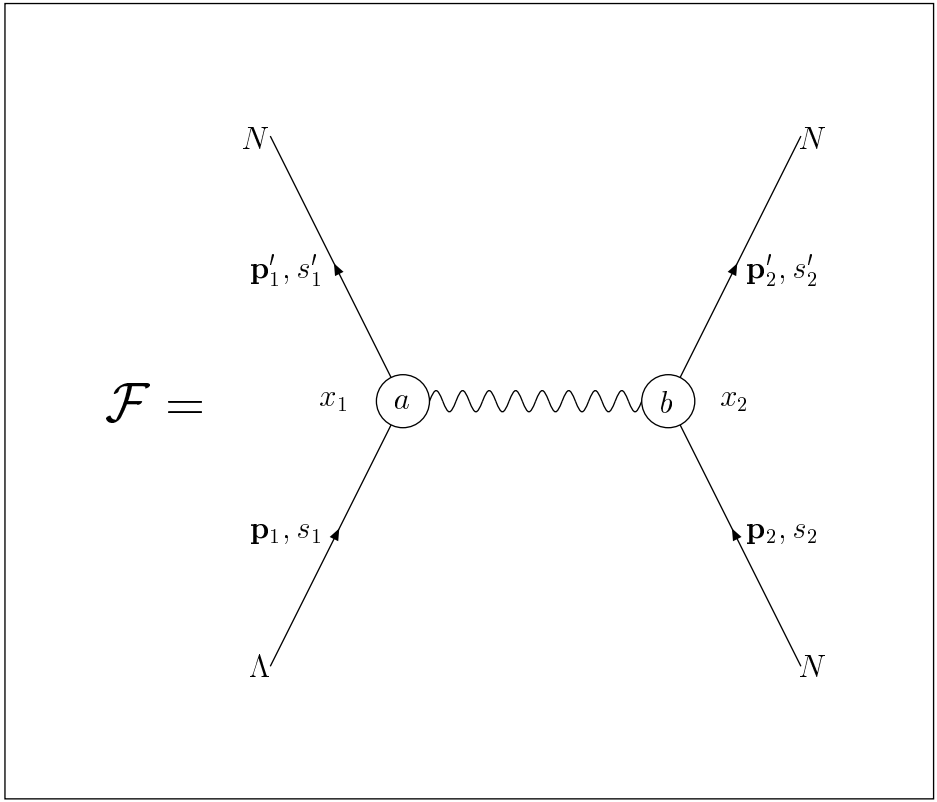}
\caption{OME Feynman amplitude in coordinate space. \label{ampc}}
\end{center}
\end{figure}
The transition rate for the nonmesonic weak decay of a hypernucleus in its 
ground state $\ket{I}$, having energy $E_I$, to a residual nucleus in any of the 
allowed final states $\bra{F}$, having energies $E_F$, and two outgoing nucleons 
is given by Fermi's golden rule,  
\begin{equation} \label{golden}
\Gamma =  2\pi \sum_{s'_1 s'_2 \, F} \, 
\int\!\int \frac{d^3 \bold{p}'_1}{(2\pi)^3} \, 
\frac{d^3 \bold{p}'_2}{(2\pi)^3} \,
\delta(E'_1 + E'_2 + E_F - E_I) \;
\left| \bra{\bold{p}'_1 s'_1\, \bold{p}'_2 s'_2\,, F} \hat{V} \ket{I} \right|^2.
\end{equation}
To construct the transition potential $\hat{V}$ in one-meson-exchange models, 
one starts from the free space Feynman amplitude depicted  in Fig.~\ref{ampc},
where $x=(t,\bold{x})$ denotes space-time coordinates, $\bold{p}$, 
momentum, and $s$, spin and, eventually, other intrinsic quantum numbers (such 
as isospin). 
In the remainder of this subsection we will consider a general situation, \ie, 
without specifying which baryons are propagating in each of the four legs, or 
which meson is being exchanged, or yet the exact nature of the couplings at the 
two vertices. 
This will be particularized to $\Lambda$-hypernuclear decay in the 
subsections that follow. 

Vertices $a$ and $b$ correspond to coupling Hamiltonians of the 
general form ($c = a$ or $b$)
\begin{equation} \label{lag}
\mathcal{H}^{c}(x) =  g_{c} \bar{\psi}(x)\, [\Gamma^{c}(\partial)\, \phi(x)]\, 
\psi(x) ,
\end{equation}
where $\psi$ and $\phi$ stand for the baryon and meson fields, respectively,
$g_{c}$ is a coupling constant  
and $\Gamma^{c}$ may contain differential operators, in which case they are 
understood to be acting on the boson field only. The Feynman rules give 
\begin{equation} \label{FF}
\mathcal{F} = (2\pi)^4\; \delta(E'_1 + E'_2 - E_1 - E_2)\, 
\delta(\bold{p}'_1 + \bold{p}'_2 - \bold{p}_1 - \bold{p}_2) \,F \;,
\end{equation}
where $E_i=\sqrt{M_i^2 + \bold{p}_i^2}\;$ ($i = 1,2$) for the incoming baryons, 
having masses $M_i$,      
and similarly (primed quantities) for the outgoing ones, and $F$ is the 
Feynman amplitude in momentum space. 
Choosing the CM frame, 
\begin{equation} \label{relmom}
-\bold{p}_1 = \bold{p}_2 = \bold{p}, \qquad 
-\bold{p}'_1 = \bold{p}'_2 = \bold{p}' ,
\end{equation}
this can be put in the form
\begin{equation} \label{F} 
iF(\bold{p}',\bold{p};s'_1 s'_2 s_1 s_2) = \chi_{s'_1}^\dagger \, 
\chi_{s'_2}^\dagger \; V(\bold{p}',\bold{p}) \; \chi_{s_1} \, \chi_{s_2}
\end{equation}
with
\begin{equation} 
V(\bold{p}',\bold{p}) = \bar{u}'_1(-\bold{p}')\,\bar{u}'_2(\bold{p}')\;
\mathcal{V}(q)\;u_1(-\bold{p})\,u_2(\bold{p}),
\end{equation}
where $u_i(\bold{p}_i) \chi_{s_i}$ and 
$\chi_{s'_i}^\dagger \,\bar{u}'_i(\bold{p}'_i)$ $(i=1,2)$ 
are the momentum eigenspinors and their conjugates for the incoming and 
outgoing 
baryons, respectively, and, denoting the meson propagator by $\mathcal{D}$,
\begin{equation} 
-i \mathcal{V}(q) = [-i g_a \Gamma^a(iq)] \,i\mathcal{D}(q)\, 
[-i g_b \Gamma^b(-iq)].
\end{equation}
We have introduced the 4-momentum transfer in the CM frame, $q = (\omega, 
\bold{q})$, with
\begin{equation} \label{trans}
\omega = \frac{1}{2} (E_1 - E'_1 + E'_2 - E_2), \qquad
\bold{q} = \bold{p}' - \bold{p}.
\end{equation}
Notice that we have directed $q$ from vertex $a$ to vertex $b$.

The nonrelativistic transition potential $\hat{V}$ is given by the 
identification 
\begin{equation} \label{Vnr}
\bra{\bold{p}'}\hat{V}\ket{\bold{p}} = V(\bold{p}',\bold{p}) ,
\end{equation}
where an expansion up to quadratic terms in momentum/mass is implied. To this 
end it is convenient to change the momentum variables to $\bold{q}$, defined in 
Eq.~\Ref{trans}, and 
\begin{equation}
\bold{Q} = \frac{m\,\bold{p}' + m'\bold{p}}{m+m'}, 
\end{equation}
where $m$ and $m'$ are the initial and final reduced masses, 
\begin{equation}
\frac{1}{m} = \frac{1}{M_1} + \frac{1}{M_2}, \qquad
\frac{1}{m'} = \frac{1}{M'_1} + \frac{1}{M'_2}.
\end{equation}
In this transformation, the following relations hold:  
\begin{equation} \label{kin}
\frac{{\bold{p}'}^2}{2m'} + \frac{\bold{p}^2}{2m} = 
\frac{1}{2} \left( \frac{1}{m'} + \frac{1}{m} \right) \bold{Q}^2 + 
\frac{\bold{q}^2}{2(m+m')} 
\end{equation} 
and 
\begin{equation} \label{phase}
\bold{p}'\cdot\bold{r}' - \bold{p}\cdot\bold{r} = 
\bold{Q}\cdot(\bold{r}'-\bold{r}) + 
\bold{q}\cdot\left(\frac{m'\bold{r}'+m\bold{r}}{m+m'}\right) ,
\end{equation}
where 
\begin{equation} \label{relative}
\bold{r} = \bold{r}_2 - \bold{r}_1 
\quad \mathrm{and} \quad
\bold{r}' = \bold{r}'_2 - \bold{r}'_1 
\end{equation}
are, respectively, the initial and final relative coordinates.

The contribution of any given meson, $i$, to Eq.~\Ref{Vnr} has the general form
\begin{equation} \label{Vi}
V_i(\bold{p}',\bold{p}) = \frac{v_i(\bold{p}',\bold{p})}{\bold{q}^2 + \mu_i^2 - 
\omega^2} \,,
\end{equation}
where $\mu_i$ is the meson mass. The nonrelativistic expansion of the numerator 
in Eq.~\Ref{Vi} poses no problem, but the denominator 
does not truly have such an expansion.\footnote{The nonrelativistic 
approximation is for the \emph{baryon} dynamics.}  
Therefore, it needs a special treatment. Recall that, 
strictly speaking, the Feynman amplitude $F$ in Eq.~\Ref{F} is defined only for 
energy-conserving transitions, \ie, for $E_1 + E_2 = E'_1 + E'_2$, and in this 
case one has, in the nonrelativistic approximation, 
\begin{equation}
\frac{\bold{p}^2}{2m} - \frac{{\bold{p}'}^2}{2m'} \cong M'_1 + M'_2 - M_1 - 
M_2.
\end{equation}
This relation together with Eq.~\Ref{kin} allow us to write, again in the 
nonrelativistic approximation, 
\begin{equation} \label{omega}
\omega \cong M_0 - \frac{\bold{q}^2}{2M_q} - \frac{\bold{Q}^2}{2M_Q} ,
\end{equation}
where we have introduced the kinematical masses $M_0$, $M_q$ and $M_Q$, given 
by 
\begin{eqnarray} 
M_0 &=& \frac{1}{2} \left[ 1 + \frac{1}{2} \left( \frac{M_1 - M_2}{M_1 + M_2} + 
\frac{M'_1 - M'_2}{M'_1 + M'_2} \right) \right] ( M_1 - M'_1 ) 
\nonumber \\ 
& & \!\!\!\!\!\!\!
{} + \frac{1}{2} \left[ 1 - \frac{1}{2} \left( \frac{M_1 - M_2}{M_1 + M_2} + 
\frac{M'_1 - M'_2}{M'_1 + M'_2} \right) \right] ( M'_2 - M_2 ) ,
\nonumber \\ & & 
\nonumber \\ 
\frac{1}{M_q} &=& \frac{1}{4} \left( \frac{M_1 - M_2}{M_1 + M_2} - \frac{M'_1 - 
M'_2}{M'_1 + M'_2} \right) \frac{1}{m + m'} \, , 
\nonumber \\  & & 
\nonumber \\ 
\frac{1}{M_Q} &=& \frac{1}{4} \left( \frac{M_1 - M_2}{M_1 + M_2} - \frac{M'_1 - 
M'_2}{M'_1 + M'_2} \right) \left( \frac{1}{m} +\frac{1}{m'} \right) .
\label{Mkin}  
\end{eqnarray}
It is clear that, whenever the absolute values 
of the differences of baryon masses are much smaller than the corresponding 
sums, as is the case, \eg, for hypernuclear decay, one can take advantage of 
the
inequalities $|M_0/M_q| \ll 1$ and $|M_0/M_Q| \ll 1$ 
to write the following approximation 
\begin{equation} \label{prop}
\frac{1}{\bold{q}^2 + \mu_i^2 - \omega^2} \cong 
\frac{1}{1 + \frac{M_0}{M_q}} \; 
\frac{1}{\bold{q}^2 + \tilde{\mu}_i^2 } 
- \frac{\frac{M_0}{M_Q}}{\left( 1 + \frac{M_0}{M_q} \right)^2} \; 
\frac{\bold{Q}^2}{(\bold{q}^2 + \tilde{\mu}_i^2)^2} \,,
\end{equation}
where we have introduced the \emph{effective meson mass} 
\begin{equation} \label{emm}
\tilde{\mu}_i = \sqrt{ \frac{\mu_i^2 - M_0^2}{1 + \frac{M_0}{M_q}} } .
\end{equation}

The net result is that the nonrelativistic approximation, as defined here, will 
reduce $V(\bold{p}',\bold{p})$ to a quadratic polynomial in $\bold{Q}$, 
\begin{equation} \label{Vmom}
V(\bold{p}',\bold{p}) \cong V^{(0)}(\bold{q}) 
+ \bold{V}^{(1)}(\bold{q})\cdot\bold{Q} 
+ \bold{Q}\cdot\boldsf{V}^{(2)}(\bold{q})\cdot\bold{Q},
\end{equation}
whose coefficients $V^{(0)}$, $\bold{V}^{(1)}$ and $\boldsf{V}^{(2)}$ are 
themselves, excluding the denominators that come from the meson propagators, 
polynomials of degree 2, 1 and 0, respectively, in $\bold{q}$. This $\bold{Q}$ 
dependence translates, in the coordinate representation, into an 
expansion in the nonlocality of the transition potential. To see this, we make 
use of Eq.~\Ref{Vnr} to write\footnote{To do this we have to assume that 
Eqs.~\Ref{Vnr} and \Ref{Vmom} hold for $\bold{p}$ and $\bold{p}'$ unrestricted, 
although $V(\bold{p}',\bold{p})$ was extracted from the Feynman amplitude $F$ 
in Eqs.~\Ref{FF} and \Ref{F}, which, being related to a T-matrix, is unambiguous  
only for energy-conserving transitions. \label{offshell}} 
\begin{eqnarray}
\bra{\bold{r}'}\hat{V}\ket{\bold{r}} &=& 
\int\frac{d^3 \bold{p}'}{(2\pi)^3} \int\frac{d^3 \bold{p}}{(2\pi)^3}\,
\braket{\bold{r}'}{\bold{p}'} 
\bra{\bold{p}'}\hat{V}\ket{\bold{p}} 
\braket{\bold{p}}{\bold{r}} 
\nonumber \\
&\equiv& \int\frac{d^3 \bold{p}'}{(2\pi)^3} \int\frac{d^3 \bold{p}}{(2\pi)^3}\,
\,e^{i\sbold{p}'\cdot\sbold{r}'}  e^{-i\sbold{p}\cdot\sbold{r}}\, 
V(\bold{p}',\bold{p}) .
\end{eqnarray}
Changing the integration variables to $(\bold{q},\bold{Q})$, making use of 
Eq.~\Ref{phase}  and truncating, for simplicity, the quadratic polynomial in 
Eq.~\Ref{Vmom} at the linear term, this gives 
\begin{equation} 
\bra{\bold{r}'}\hat{V}\ket{\bold{r}} \cong 
\left[ V^{(0)}(\bold{r}') 
- i \bold{V}^{(1)}\left({\textstyle 
\frac{m'\sbold{r}'+m\sbold{r}}{m+m'}}\right)
\cdot\bold{\nabla}' \right] \delta(\bold{r}' - \bold{r}) ,
\end{equation}
where 
\begin{equation}\label{V0r}
V^{(0)}(\bold{r}) = \int \frac{d^3 \bold{q}}{(2\pi)^3}\, 
\,e^{i\sbold{q}\cdot\sbold{r}}\, V^{(0)}(\bold{q}) 
\end{equation}
and
\begin{equation} \label{V1r}
\bold{V}^{(1)}(\bold{r}) = \int \frac{d^3 \bold{q}}{(2\pi)^3}\, 
\,e^{i\sbold{q}\cdot\sbold{r}}\, \bold{V}^{(1)}(\bold{q}). 
\end{equation}
This means that for any state of relative motion, $\Psi$, 
\begin{equation}
\bra{\bold{r}}\hat{V}\ket{\Psi} = \hat{V}(\bold{r}) \Psi(\bold{r}),
\end{equation}
with the \emph{transition potential in coordinate space}, $\hat{V}(\bold{r})$,
given, as an operator in wave-function space, by
\begin{equation}
\hat{V}(\bold{r}) = V^{(0)}(\bold{r}) + \hat{V}^{(1)}(\bold{r})
\end{equation}
where $V^{(0)}(\bold{r})$ is the \emph{local potential}, and the differential 
operator
\begin{equation} \label{Vnl}
\hat{V}^{(1)}(\bold{r}) = 
- i \frac{m}{m+m'}\, \left( \bold{\nabla} \cdot \bold{V}^{(1)}(\bold{r}) \right)
- i \bold{V}^{(1)}(\bold{r})\cdot\bold{\nabla} ,
\end{equation}
its \emph{first-order nonlocality correction}\footnote{Notice that, despite its 
name, this has a local piece, namely, the first term in Eq.~\Ref{Vnl}, where 
$\left( \bold{\nabla} \cdot \bold{V}^{(1)}(\bold{r}) \right) \equiv  
\mathrm{div} \bold{V}^{(1)}(\bold{r})$.}.
For our purposes here it will be sufficient to stop at this order, and  we will 
not consider the second-order corrections, that would come from the last term 
in Eq.~\Ref{Vmom}.

Before closing this subsection, let us make a brief comment on our treatment of 
the $\omega^2$ term in the meson propagator in Eq.~\Ref{Vi}, more specifically, 
on the expression we use for the time component, $\omega$, of the 4-momentum 
transfer in Eq.~\Ref{trans}. We are extracting the transition potential from the 
Feynman diagram in Fig.~\ref{ampc}, in which the baryons are on their mass 
shells. However, as already alluded to in footnote~\ref{offshell}, we need to 
extend these OME amplitudes to the off-shell region, for which there is no 
unique procedure. In the meson-exchange theory for the strong NN force, this is 
done by treating the two interacting particles through the Bethe-Salpeter 
equation \cite{Sa51}, and this ambiguity appears in the choice 
of which one of its various tridimensional reductions to use. This issue, which 
is related to meson retardation effects, has been much discussed in the past 
\cite{Sc72}--\cite{Gr92}, but remains  unsettled.
We have followed the general philosophy of Machleidt and collaborators 
\cite{Ma87}--\cite{Br90}, 
to the effect that, for the case of similar masses, the best choice is to use 
tridimensional reductions that 
treat the two particles symmetrically, like the Blankenbecler-Sugar \cite{Bl66} 
or the Thompson \cite{Th70} equations. 
One, then, puts the two interacting particles equally off-shell in the CM frame, 
and fixes the time-components of the relative 4-momenta in the initial and final 
two-particle propagators at the values $p_0 = \frac{1}{2} (E_2 - E_1)$ 
and $p'_0 = \frac{1}{2} (E'_2 - E'_1)$, respectively. 
(See, for instance, Eq. (2.28) in Ref.~\cite{Si84}, remembering our convention 
in Eq.~\Ref{relmom}.) This leads directly to our expression for $\omega$ in 
Eq.~\Ref{trans}. Notice that, for strictly equal masses, this gives 
$\omega = 0$, being, therefore, equivalent, in this case, to the instantaneous 
approximation. 
Our choice for $\omega$ in Eq.~\Ref{trans} leads, in the nonrelativistic 
approximation, to Eq.~\Ref{omega} and, consequently, to the expansion 
of the meson propagators in Eq.~\Ref{prop}. We are confident that this is 
appropriate for processes not too far off the energy-shell. In other situations 
that might occur, for instance, in a fully microscopic treatment of short-range 
correlations, this point should be reexamined.

%
%
\subsection{Kinematical effects} \label{Kin}

 In computing the OME Feynman amplitudes contributing to the strong $NN$ force 
it is standard practice \cite{Ma89} to avoid the kinematical
complications due to the difference between the neutron and proton masses 
by setting 
\begin{equation}\label{nep}
M_n , M_p \; \rightarrow \; M \equiv (M_n + M_p)/2,
\end{equation} 
which can be justified by the small value of the ratio
\begin{equation}\label{ndp}
\frac{M_n - M_p}{M} = 0.0014.
\end{equation}
The analogous practice is followed in the calculation of the transition 
potentials for the weak decay of $\Lambda$-hypernuclei \cite{Pa97}. In this 
case, however, one equally sidesteps the lambda-nucleon mass difference, by 
setting, at the vetex where the $\Lambda$ decays, 
\begin{equation}\label{len}
M_\Lambda , M \; \rightarrow \; \bar{M} \equiv (M_\Lambda + M)/2,
\end{equation}
despite the fact that the corresponding ratio,
\begin{equation}\label{ldn}
\frac{M_\Lambda - M}{\bar{M}} = 0.17,
\end{equation}
is nowhere as small.  

Undoubtedly, this approximation considerably simplifies 
the calculations. However, in view of the nonnegligible value of the ratio 
\Ref{ldn}, it seems appropriate to investigate the effects of the latter 
approximation. To this end, we examine below, for each meson in the pseudoscalar 
and vector octets, the expression 
for the nonrelativistic OME transition potential obtained by 
accepting the approximation in Eq.\Ref{nep}, but not that in Eq.\Ref{len}. 
This gives for the kinematical masses \Ref{Mkin}  
\[  
M_0 = \frac{1}{4} \left( \frac{M_\Lambda - M}{M_\Lambda + M} \right) 
\left(3M_\Lambda + M \right) = 92.18 \mbox{ MeV} \,, 
\]  
\[  
\frac{1}{M_q} = \frac{1}{2} \left( \frac{M_\Lambda - M}{3M_\Lambda + M} \right)
\frac{1}{M} =  2.196 \times 10^{-5} \mbox{ MeV}^{-1} \,,
\]  
\begin{equation}\label{mkin}
\frac{1}{M_Q} = \frac{1}{4} \left( \frac{M_\Lambda - M}{M_\Lambda + M} \right) 
\left( \frac{3 M_\Lambda + M}{M_\Lambda M} \right) 
= 8.800 \times 10^{-5} \mbox{ MeV}^{-1}, 
\end{equation}
where we have used \cite{Ba96} $M = 938.92 \mbox{~MeV}$ and $M_\Lambda = 
1115.68 
\mbox{ MeV}$ \,.
The approximation \Ref{len} would have set $M_0$, $1/M_q$ and $1/M_Q$ to zero. 

\subsection{Contributions of nonstrange mesons}\label{nstr}

For the nonstrange mesons, we have, acting respectively at the vertices $a$ and 
$b$ in Fig.~\ref{ampc}, weak ($W$) and strong ($S$) coupling Hamiltonians that 
we take to be the same as those in Ref.~\cite{Pa97}. 
For the pion they are  
\begin{equation}
\mathcal{H}^W_{\Lambda N \pi} = i G_F \, \mu_\pi^2 \, \bar{\psi}_N ( A_\pi + 
B_\pi \gamma_5 )\, \bold{\phi}^{(\pi)}\cdot\bold{\tau} \, \spurion\, 
\psi_\Lambda,
\end{equation}
\begin{equation}
\mathcal{H}^S_{N N \pi} = i g_{NN\pi}\, \bar{\psi}_N \gamma_5\,  
\bold{\phi}^{(\pi)}\cdot\bold{\tau}\, \psi_N ,
\end{equation}
where $G_F \, \mu_\pi^2 = 2.21 \times 10^{-7}$ is the Fermi weak 
coupling-constant, $A_\pi$ and $B_\pi$ are fitted to the 
free $\Lambda$ decay, and $g_{NN\pi}$ is taken from OME models 
for the strong $NN$ force.
The isospurion $\spurion$ is used to enforce the $\Delta T = \frac{1}{2}$ 
rule for isospin violation, observed in the free $\Lambda$ decay \cite{Pa97}. 
For the rho meson, we have
\begin{equation}
\mathcal{H}^W_{\Lambda N \rho} =  G_F\, \mu_\pi^2\, \bar{\psi}_N 
\left[ 
\left( A_\rho \gamma^\nu \gamma_5 
+ B^V_\rho\, \gamma^\nu 
+ B^T_\rho\, \frac{\sigma^{\mu\nu}\partial_\mu}{2\bar{M}} \right)\, 
\bold{\phi}^{(\rho)}_{\,\nu}\cdot\bold{\tau} \right]\, 
\spurion\, \psi_\Lambda,
\end{equation}
\begin{equation}
\mathcal{H}^S_{N N \rho} =  \bar{\psi}_N 
\left[ \left( g^V_{NN\rho}\, \gamma^\nu 
+ g^T_{NN\rho}\, \frac{\sigma^{\mu\nu}\partial_\mu}{2M} \right)\, 
\bold{\phi}^{(\rho)}_{\,\nu}\cdot\bold{\tau} \right]\, \psi_N \,. 
\end{equation}
The corresponding Hamiltonians for the $\eta$ and $\omega$ are completely 
analogous to those of the $\pi$ and $\rho$, respectively, if one takes into 
consideration their isoscalar nature.  

The weak couplings of the heavier mesons are theoretically inferred from those 
of the pion through unitary-symmetry arguments and other relationships. 
The strong ones are again taken from OME models for the nuclear force. 
We will follow the parametrization 
adopted in Ref.~\cite{Pa97}, where further details can be found on this matter. 
For definiteness, the numerical values are reproduced in Table~\ref{nonstrange}.
\settoheight{\thisheight}{ $ \displaystyle B^T_\rho $ }
\begin{table}[!ht]   
\vspace{.5\baselineskip}
\begin{center}
\caption{\label{nonstrange}
Coupling constants, masses ($\mu_i$) and cutoff parameters ($\Lambda_i$) for the 
nonstrange mesons. The weak couplings are in units of $G_F\,\mu_\pi^2$.  Adapted 
from Ref.~\cite{Pa97}.}
\vspace{.5\baselineskip}

\begin{tabular*}{\textwidth}{@{\extracolsep{\fill}}|clllcc|} 
\hline
\hline 
Meson & \multicolumn{3}{c}{Coupling Constants} & $\mu_i$ & $\Lambda_i$ \\
\cline{2-4}
$i$ & \multicolumn{2}{c}{Weak} & \multicolumn{1}{c}{Strong} & [MeV] & [GeV] \\
\cline{2-3}
 & \multicolumn{1}{c}{$PV$} & \multicolumn{1}{c}{$PC$} & & & \\ 
\hline
& & & & & \\
$\pi$ &  
$A_\pi = 1.05$ & $B_\pi = -7.15$ & $g_{NN\pi} = 13.3$ 
& 140.0 & 1.30 \\
& & & & & \\
$\eta$ & 
$A_\eta = 1.80$ & $B_\eta = -14.3$ & $g_{NN\eta} = 6.40$ 
& 548.6 & 1.30 \\
& & & & & \\ 
$\rho$ &
$A_\rho = 1.09$ & $B^V_\rho = -3.50$ & $g^V_{NN\rho} = 3.16$ 
& 775.0 & 1.40 \\
&\rule{0cm}{1.5\thisheight} & $B^T_\rho = -6.11$ & $g^T_{NN\rho} = 13.3$ 
& & \\
& & & & & \\
$\omega$ & 
$A_\omega = -1.33$ & $B^V_\omega = -3.69$ & $g^V_{NN\omega} = 10.5$ 
& 783.4 & 1.50 \\
&\rule{0cm}{1.5\thisheight} & $B^T_\omega = -8.04$ & $g^T_{NN\omega} = 3.22$ 
& & \\
& & & & & \\
\hline\hline 
\end{tabular*} 
\end{center}
\end{table}

\subsubsection{One pion exchange} \label{pion} 

The local nonrelativistic one-pion-exchange transition potential is given in 
momentum space by
\begin{equation}\label{vpi}
V^{(0)}_\pi(\bold{q}) = - \left(1+\frac{M_0}{M_q}\right)^{-1} 
G_F\,\mu_\pi^2 \,
\frac{g_{NN\pi}}{2M} \, \bold{\tau}_1 \cdot \bold{\tau}_2 \,  
\left( A_\pi + \frac{B_\pi}{2\check{M}}\, 
\bold{\sigma}_1 \cdot \bold{q} \right) \; 
\frac{\bold{\sigma}_2 \cdot \bold{q}}{\bold{q}^2 + \tilde{\mu}_\pi^2} \,,
\end{equation}
where  
\begin{equation}
\check{M} = \frac{M + 3M_\Lambda}{3M + M_\Lambda}\, M .
\end{equation}
Comparing Eq.\Ref{vpi} with the result that would have been obtained under 
approximation \Ref{len}, namely, \cite[\emph{Eq.(24)}]{Pa97}
\begin{equation}
\bar{V}^{(0)}_\pi(\bold{q}) = - G_F\,\mu_\pi^2 \,
\frac{g_{NN\pi}}{2M} \, \bold{\tau}_1 \cdot \bold{\tau}_2 \,  
\left( A_\pi + \frac{B_\pi}{2\bar{M}}\, 
\bold{\sigma}_1 \cdot \bold{q} \right) \; 
\frac{\bold{\sigma}_2 \cdot \bold{q}}{\bold{q}^2 + \mu_\pi^2} \,,
\end{equation}
it is possible to estimate the relative size of the effects of the more accurate 
treatment of the kinematics, adopted here, from the following correction 
factors:
\begin{equation}\label{cf12}
(1 + M_0/M_q)^{-1} = 0.998, \qquad
\check{M}/\bar{M} = 0.996
\end{equation}
and
\begin{equation} \label{cf3}
\tilde{\mu}_\pi/\mu_\pi = 0.752.   
\end{equation}
If each of these values were equal to unity, there would be no effect at all. 
Apparently, the situation is not much different from this, except in the last 
case, which will have a noticeable effect 
since it increases by $\sim 35\%$ the range of the 
corresponding contribution to the transition potential.

When changing to the coordinate representation through Eqs.~\Ref{V0r} or 
\Ref{V1r}, the shape functions 
\begin{eqnarray}
f_C(r,\mu) &=& \int \frac{d^3 \bold{q}}{(2\pi)^3}\; 
e^{i\sbold{q}\cdot\sbold{r}} \, \frac{1}{\bold{q}^2 + \mu^2} 
\;\; = \;\; 
\frac{e^{-\mu r}}{4 \pi r} \,,
\nonumber 
\\
f_V(r,\mu) &=& - \frac{\partial}{\partial r} f_C(r,\mu) \;\; = \;\;
\mu \left( 1 + \frac{1}{\mu r} \right) f_C(r,\mu) \,,
\nonumber 
\\
f_S(r,\mu) &=& \frac{1}{3} \left[ \mu^2 \, f_C(r,\mu) 
- \delta(\bold{r}) \right] \,,
\nonumber 
\\
f_T(r,\mu) &=& \frac{1}{3}\,\mu^2
\left[ 1 + \frac{3}{\mu r} + \frac{3}{(\mu r)^2} \right]
f_C(r,\mu) 
\label{fCVST}
\end{eqnarray}
naturally arise, accordingly as the numerators in the Fourier transforms are, 
respectively, a constant, a vector, a scalar or a tensor built, at most 
quadratically, from $\bold{q}$. In terms of these, we get, for the potential 
\Ref{vpi} in coordinate space,
%
\settoheight{\thisheight}{ $ \displaystyle \frac{B_\pi}{2\check{M}} $ }
\begin{eqnarray}
V^{(0)}_\pi(\bold{r}) &=&  \left(1+\frac{M_0}{M_q}\right)^{-1} 
G_F\,\mu_\pi^2 \;
\frac{g_{NN\pi}}{2M} \; \bold{\tau}_1 \cdot \bold{\tau}_2 \; 
\left[ 
\rule{0cm}{\thisheight}  
- i A_\pi\, f_V(r,\tilde{\mu}_\pi)\, \bold{\sigma}_2 \cdot \hat{\bold{r}}  
\right.
\nonumber \\ & & \left. {}
+ \frac{B_\pi}{2\check{M}}\, f_S(r,\tilde{\mu}_\pi)\, 
\bold{\sigma}_1 \cdot \bold{\sigma}_2 
+ \frac{B_\pi}{2\check{M}}\, f_T(r,\tilde{\mu}_\pi)\, S_{12}(\hat{\bold{r}})  
\right] \, ,
\label{fullV0pic} 
\end{eqnarray}
where $\hat{\bold{r}} = \bold{r}/r$ and $S_{12}(\hat{\bold{r}}) = 
3(\bold{\sigma}_1\cdot\hat{\bold{r}})(\bold{\sigma}_2\cdot\hat{\bold{r}}) - 
\bold{\sigma}_1\cdot\bold{\sigma}_2$. Under approximation \Ref{len}, this would 
reduce to 
\begin{eqnarray}
\bar{V}^{(0)}_\pi(\bold{r}) &=&  G_F\,\mu_\pi^2 \;
\frac{g_{NN\pi}}{2M} \; \bold{\tau}_1 \cdot \bold{\tau}_2 \; 
\left[ \rule{0cm}{\thisheight} 
- i A_\pi\, f_V(r,\mu_\pi)\, \bold{\sigma}_2 \cdot \hat{\bold{r}} 
\right.
\nonumber \\ & & \left. {}
+ \frac{B_\pi}{2\bar{M}}\, f_S(r,\mu_\pi)\, 
\bold{\sigma}_1 \cdot \bold{\sigma}_2  
+ \frac{B_\pi}{2\bar{M}}\, f_T(r,\mu_\pi) S_{12}(\hat{\bold{r}}) 
\right] . 
\label{avV0pic}
\end{eqnarray}
%


The first-order nonlocality coefficient in momentum space, appearing in 
Eq.~\Ref{Vmom}, is given, for the pion, by
\begin{equation} \label{V1pi}
\bold{V}^{(1)}_\pi(\bold{q}) = - \left(1+\frac{M_0}{M_q}\right)^{-1} 
G_F\,\mu_\pi^2 \,
\frac{g_{NN\pi}}{2M} \, \bold{\tau}_1 \cdot \bold{\tau}_2 \,  
\left( \frac{B_\pi}{2\grave{M}}\,\bold{\sigma}_1  \right) \; 
\frac{\bold{\sigma}_2 \cdot \bold{q}}{\bold{q}^2 + \tilde{\mu}_\pi^2} \,,
\end{equation}
where 
\begin{equation} \label{graveM}
\frac{1}{\grave{M}} = \frac{1}{M} - \frac{1}{M_\Lambda} \,.
\end{equation}
Changing to the coordinate representation according to Eq.~\Ref{V1r}, we get 
\begin{equation}
\bold{V}^{(1)}_\pi(\bold{r}) = - i \left(1+\frac{M_0}{M_q}\right)^{-1} 
G_F\,\mu_\pi^2 \,
\frac{g_{NN\pi}}{2M} \, \frac{B_\pi}{2\grave{M}}\,
\bold{\tau}_1 \cdot \bold{\tau}_2 \,  
f_V(r,\tilde{\mu}_\pi) \, 
( \bold{\sigma}_2 \cdot \hat{\bold{r}} ) \; \bold{\sigma}_1 
\end{equation}
and introducing this into Eq.~\Ref{Vnl} yields for the first-order nonlocality 
correction 
%
%
\settoheight{\thisheight}{ $ \displaystyle 
\frac{2 M_\Lambda}{3 M_\Lambda + M} $ }
\begin{eqnarray}
\hat{V}^{(1)}_\pi(\bold{r}) &=&  \left(1+\frac{M_0}{M_q}\right)^{-1} 
G_F\,\mu_\pi^2 \;
\frac{g_{NN\pi}}{2M} \; 
\frac{B_\pi}{2\grave{M}}\;
\bold{\tau}_1 \cdot \bold{\tau}_2 \; \times
\nonumber \\  & & 
\left\{ 
\frac{2 M_\Lambda}{3 M_\Lambda + M} 
\left[  
f_S(r,\tilde{\mu}_\pi) \,  \bold{\sigma}_1 \cdot \bold{\sigma}_2 
+ f_T(r,\tilde{\mu}_\pi) \;  S_{12}(\hat{\bold{r}}) \right] \right.
\nonumber \\ & & \left.  {}
- f_V(r,\tilde{\mu}_\pi) \, 
( \bold{\sigma}_2 \cdot \hat{\bold{r}} ) \; 
( \bold{\sigma}_1 \cdot \bold{\nabla} ) 
\rule{0cm}{\thisheight} 
\right\} .
\label{fullV1pic} 
\end{eqnarray}
The mass averaging approximation \Ref{len} would set $1/\grave{M}$ to zero. 
Therefore, $\bar{\bold{V}}^{(1)}_\pi(\bold{q}) = 0$ and there would be no 
first-order nonlocality correction for the pion under this approximation, \ie, 
\begin{equation} \label{avV1pic} 
\hat{\bar{V}}^{(1)}_\pi(\bold{r}) = 0 \,.
\end{equation}

\subsubsection{One rho exchange} \label{rho} 

The one-rho-exchange contribution to the local nonrelativistic transition 
potential in momentum space is 
\begin{eqnarray}
V^{(0)}_\rho(\bold{q}) &=& \left(1+\frac{M_0}{M_q}\right)^{-1}\, 
G_F\,\mu_\pi^2\; \bold{\tau}_1\cdot\bold{\tau}_2 
\Bigm[ K^1_\rho - K^2_\rho\, \bold{q}^2
\nonumber \\
& & {} - K^3_\rho\, (\bold{\sigma}_1 \times \bold{q}) \cdot 
        (\bold{\sigma}_2 \times \bold{q})   
\nonumber \\
& & {} - i\,K^4_\rho\, (\bold{\sigma}_1 \times \bold{\sigma}_2) \cdot \bold{q}   
 + K^5_\rho \, (\bold{\sigma}_1 \cdot \bold{q}) \Bigm]  
\frac{1}{\bold{q}^2 + \tilde{\mu}_\rho^2} \,, 
\label{vrho}              
\end{eqnarray}
where, for notational convenience, we have introduced the coefficients
\begin{eqnarray}
K^1_\rho &=& B^V_\rho \, g^V_{NN\rho} \,,
\nonumber \\
K^2_\rho &=& B^V_\rho \, g^V_{NN\rho} \,  
                 \left[ \left( \frac{1}{4M} \right)^2 
                      + \left( \frac{1}{4\check{M}} \right)^2  
                      + \frac{1}{2} \left( \frac{1}{M_q} \right)^2  \right] 
\nonumber \\       
& & {} + \left( \frac{B^V_\rho}{2M}\,\frac{g^T_{NN\rho}}{2M} 
       + \frac{B^T_\rho}{2\bar{M}}\,\frac{g^V_{NN\rho}}{2\check{M}} \right)\,
         \left( 1 + \frac{M_0}{M_q} \right)
\nonumber \\          
& & {} + \frac{B^T_\rho}{2\bar{M}}\,\frac{g^T_{NN\rho}}{2M} \, 
         \left( \frac{M_0^2}{4M\check{M}} \right) ,
\nonumber \\
K^3_\rho &=& \left[ \frac{B^V_\rho}{2\check{M}} +  
\frac{B^T_\rho}{2\bar{M}}\,\left( 1 + \frac{M_0}{M_q} \right) \right]   
\left[ \frac{g^V_{NN\rho}}{2M} + \frac{g^T_{NN\rho}}{2M}\,
\left( 1 + \frac{M_0}{M_q} \right) \right] ,
\nonumber \\
K^4_\rho &=&  A_\rho \,
\left[ \frac{g^V_{NN\rho}}{2M} + \frac{g^T_{NN\rho}}{2M}
\left( 1 + \frac{M_0}{M_q} \right) \right] ,
\nonumber \\
K^5_\rho &=& A_\rho \, \frac{g^T_{NN\rho}}{2M} 
\left(\frac{M_0}{2M}\right) .
\label{Krho}          
\end{eqnarray}

The corresponding potential under approximation \Ref{len}, 
$\bar{V}^{(0)}_\rho(\bold{q})$, can be obtained from Eq.~\Ref{vrho} through the 
substitutions:\footnote{This agrees with   
Eq. (34) of Ref. \cite{Pa97}, except for a wrong sign 
($A_\rho \rightarrow - A_\rho$) and an omitted term 
($\propto \bold{q}^2$).} 
\begin{equation} \label{subs}
V^{(0)}_\rho \rightarrow \bar{V}^{(0)}_\rho\,,
\qquad K^j_\rho \rightarrow \bar{K}^j_\rho \;\; (j=1 \mbox{--} 5)\,,
\qquad M_0/M_q \rightarrow 0\,, 
\qquad \tilde{\mu}_\rho \rightarrow \mu_\rho\,,
\end{equation}
with
\begin{eqnarray}
\bar{K}^1_\rho &=& B^V_\rho \, g^V_{NN\rho} \,,
\nonumber \\
\bar{K}^2_\rho &=& B^V_\rho \, g^V_{NN\rho} \,  
\left[ \left( \frac{1}{4M} \right)^2 +
       \left( \frac{1}{4\bar{M}} \right)^2  \right] + 
       \frac{B^V_\rho}{2M}\,\frac{g^T_{NN\rho}}{2M} +
       \frac{B^T_\rho}{2\bar{M}}\,\frac{g^V_{NN\rho}}{2\bar{M}} \,,
\nonumber \\
\bar{K}^3_\rho &=& \left(\frac{B^V_\rho + B^T_\rho}{2\bar{M}}\right) \,
            \left(\frac{g^V_{NN\rho} + g^T_{NN\rho}}{2M}\right) \,,
\nonumber \\
\bar{K}^4_\rho &=& A_\rho \, 
              \left(\frac{g^V_{NN\rho} + g^T_{NN\rho}}{2M}\right) \,, 
\nonumber \\
\bar{K}^5_\rho &=& 0 \,.
\label{bKrho}          
\end{eqnarray}

Let us now compare Eqs.~\Ref{Krho} with the corresponding 
expressions under approximation \Ref{len}, namely, Eqs.~\Ref{bKrho}. 
Firstly, we notice that the two correction factors in Eq.~\Ref{cf12} as well as  
\[ 
\tilde{\mu}_\rho/\mu_\rho = 0.992 
\]
are very close to unity. Secondly, we also notice that the relative values of 
the remaining correction terms can be estimated from
\[  
\left.
\frac{1}{2} \left( \frac{1}{M_q} \right)^2
\right/ 
\left[
\left( \frac{1}{4M} \right)^2 + \left( \frac{1}{4\check{M}} \right)^2 
\right]
= 1.85\times 10^{-3}, 
\] 
\[
\frac{M_0^2}{4M\check{M}} = 2.21\times 10^{-3} , \qquad
\frac{M_0}{2M} = 4.91\times 10^{-2}
\]
and are, therefore, considerably less than unity. We, thus, expect  
that, as far as the local contributions are concerned, only very small 
corrections will result from the more accurate treatment of the kinematics in 
the present case.  

Making use of Eq.~\Ref{V0r}, we get, for  the potential \Ref{vrho} in coordinate 
space, 
\begin{eqnarray} 
V^{(0)}_{\rho}(\bold{r}) &=& \left(1+\frac{M_0}{M_q}\right)^{-1} 
G_F\,\mu_\pi^2\;\bold{\tau}_1\cdot\bold{\tau}_2\; 
\left\{ K^1_\rho\, f_C(r,\tilde{\mu}_\rho) 
+ 3 K^2_\rho\, f_S(r,\tilde{\mu}_\rho) 
\right. 
\nonumber \\
& & {} + 2 K^3_\rho\, f_S(r,\tilde{\mu}_\rho)\, 
\bold{\sigma}_1 \cdot \bold{\sigma}_2
- K^3_\rho\, f_T(r,\tilde{\mu}_\rho)\, S_{12}(\hat{\bold{r}}) 
\nonumber \\ 
& & \left. {} + f_V(r,\tilde{\mu}_\rho)  
\left[ K^4_\rho \, (\bold{\sigma}_1 \times \bold{\sigma}_2)  
+ i K^5_\rho \, \bold{\sigma}_1 \right] \cdot \hat{\bold{r}} \right\} \, .  
\label{vrhoc}
\end{eqnarray}
Once more, the corresponding potential under approximation \Ref{len} can be 
obtained from the above equation by means of the substitutions \Ref{subs}.

For the rho meson, the coefficient of the linear term in Eq.~\Ref{Vmom} is given 
by
\begin{eqnarray} 
\bold{V}^{(1)}_\rho(\bold{q}) &=& - \left(1+\frac{M_0}{M_q}\right)^{-1}
G_F\,\mu_\pi^2\, \bold{\tau}_1 \cdot \bold{\tau}_2 \times 
\nonumber \\ & & 
\left[ K^6_\rho\, \bold{q} 
      - i K^7_\rho\, \bold{\sigma}_1 \times \bold{q}  
      - i K^8_\rho\, \bold{\sigma}_2 \times \bold{q} \right.
\nonumber \\ & & \left. {} 
   - K^9_\rho\, (\bold{\sigma}_1\times\bold{q}) \times \bold{\sigma}_2 
+ K^{10}_\rho\, (\bold{\sigma}_2\times\bold{q}) \times \bold{\sigma}_1 \right.   
\nonumber \\ & & \left. {}     
      + K^{11}_\rho\, \bold{\sigma}_1 
    - i K^{12}_\rho\, \bold{\sigma}_1\times\bold{\sigma}_2  \right] 
\frac{1}{\bold{q}^2 + \tilde{\mu}_\rho^2} \,,
\label{V1m}
\end{eqnarray}
where 
\begin{eqnarray}
K^6_\rho &=& B^V_\rho\, g^V_{NN\rho}\, \frac{1}{16\grave{M}} \, 
\left( \frac{3}{\check{M}} - \frac{1}{M} \right) 
+ B^V_\rho\, \frac{g^T_{NN\rho}}{2M}\, \frac{M_0}{4M\acute{M}}  
\nonumber \\ & & {} 
+ \frac{B^T_\rho}{2\bar{M}}\, g^V_{NN\rho}\, 
\left[ \frac{1}{2\grave{M}}\, \left(1+\frac{M_0}{M_q}\right) - 
\frac{M_0}{2M\check{M}} \right]
+ \frac{B^T_\rho}{2\bar{M}}\, \frac{g^T_{NN\rho}}{2M}\, 
\frac{M_0^2}{4M\grave{M}} \,,
\nonumber \\
K^7_\rho &=& B^V_\rho\, g^V_{NN\rho}\, 
\left[ \frac{1}{8\check{M}}\, \left( \frac{2}{M} + \frac{1}{\acute{M}} \right) 
+  \frac{1}{8M\acute{M}} \right]  
+ B^V_\rho\, \frac{g^T_{NN\rho}}{2M}\, \frac{M_0}{4M\grave{M}}  
\nonumber \\ & & {} 
+ \frac{B^T_\rho}{2\bar{M}}\, g^V_{NN\rho}\, 
\left[ \frac{1}{2}\, \left( \frac{2}{M} + \frac{1}{\acute{M}} \right)\, 
\left(1+\frac{M_0}{M_q}\right) \right]
+ \frac{B^T_\rho}{2\bar{M}}\, \frac{g^T_{NN\rho}}{2M}\, 
\frac{M_0^2}{4M\acute{M}} \,,
\nonumber \\
K^8_\rho &=& B^V_\rho\, g^V_{NN\rho}\, 
\frac{1}{4M}\, \left( \frac{1}{M} + \frac{1}{\acute{M}} \right) 
+ B^V_\rho\, \frac{g^T_{NN\rho}}{2M}\, 
\left[ \frac{1}{2}\, \left( \frac{2}{M} + \frac{1}{\acute{M}} \right)\, 
\left(1+\frac{M_0}{M_q}\right) \right] 
\nonumber \\ & & {} 
+ \frac{B^T_\rho}{2\bar{M}}\, g^V_{NN\rho}\, \frac{M_0}{4M\grave{M}}
+ \frac{B^T_\rho}{2\bar{M}}\, \frac{g^T_{NN\rho}}{2M}\, \frac{M_0}{2}\,
\left[ \frac{1}{\grave{M}}\, \left(1+\frac{M_0}{M_q}\right) 
+ \frac{M_0}{M\check{M}} \right] \,,
\nonumber \\
K^9_\rho &=&  B^V_\rho\, \frac{g^T_{NN\rho}}{2M}\, 
\frac{M_0}{2M\check{M}}
+ \frac{B^T_\rho}{2\bar{M}}\, \frac{g^T_{NN\rho}}{2M}\, \frac{M_0}{M}\,
\left(1+\frac{M_0}{M_q}\right) \,,
\nonumber \\
K^{10}_\rho &=& B^V_\rho\, g^V_{NN\rho}\, 
\frac{1}{4M\grave{M}}\,  
+ B^V_\rho\, \frac{g^T_{NN\rho}}{2M}\, 
\frac{1}{2\grave{M}}\, \left(1+\frac{M_0}{M_q}\right) 
\nonumber \\ & & {} 
+ \frac{B^T_\rho}{2\bar{M}}\, g^V_{NN\rho}\, \frac{M_0}{4M\acute{M}}
+ \frac{B^T_\rho}{2\bar{M}}\, \frac{g^T_{NN\rho}}{2M}\, 
\frac{M_0}{2\acute{M}}\, \left(1+\frac{M_0}{M_q}\right) \,,
\nonumber \\
K^{11}_\rho &=&  A_\rho\, g^V_{NN\rho}\, 
\left[ \frac{1}{2}\, 
\left( \frac{2}{M} + \frac{1}{\acute{M}} \right) \right] \,,
\nonumber \\
K^{12}_\rho &=&  A_\rho\, \frac{g^T_{NN\rho}}{2M}\, 
\frac{M_0}{M} \,,
\end{eqnarray}
with
\begin{equation}\label{acuteM}
\frac{1}{\acute{M}} = \frac{1}{M} + \frac{1}{M_\Lambda} \,.
\end{equation}
To get the first-order nonlocality correction  $\hat{V}^{(1)}_\rho(\bold{r})$, 
we need first to change \Ref{V1m} to the coordinate representation, according to 
Eq.~\Ref{V1r}. This gives
\settoheight{\thisheight}{$ \displaystyle \frac{f_V(r,\mu_\rho)}{r} $}
\begin{eqnarray} 
\bold{V}^{(1)}_\rho(\bold{r}) &=& - \left(1+\frac{M_0}{M_q}\right)^{-1}
G_F\,\mu_\pi^2\, \bold{\tau}_1 \cdot \bold{\tau}_2 \times 
\nonumber \\ & & 
\left\{ \frac{f_V(r,\tilde{\mu}_\rho)}{r}\,
\left[ i K^6_\rho\, \bold{r} 
       + K^7_\rho\, \bold{\sigma}_1 \times \bold{r}  
       + K^8_\rho\, \bold{\sigma}_2 \times \bold{r} \right. \right.
\nonumber \\ & & \left. {} 
   - i K^9_\rho\, (\bold{\sigma}_1\times\bold{r}) \times \bold{\sigma}_2 
+ i K^{10}_\rho\, (\bold{\sigma}_2\times\bold{r}) \times \bold{\sigma}_1 \right]   
\nonumber \\ & & \left. {} 
+ f_C(r,\tilde{\mu}_\rho)\, \left[     
        K^{11}_\rho\, \bold{\sigma}_1 
    - i K^{12}_\rho\, \bold{\sigma}_1\times\bold{\sigma}_2  \right] 
\rule{0cm}{\thisheight} \right\} \,. 
\label{V1c}
\end{eqnarray}
Introducing \Ref{V1c} into Eq.~\Ref{Vnl} and noticing that 
$-i\bold{\sigma}\times\bold{r}\cdot\bold{\nabla} = \bold{\sigma}\cdot\bold{l}$, 
where $\bold{l} = -i\bold{r}\times\bold{\nabla}$ is the relative orbital 
angular momentum, we obtain, finally,
\begin{eqnarray} 
\hat{V}^{(1)}_\rho(\bold{r}) &=&  \left(1+\frac{M_0}{M_q}\right)^{-1}
G_F\,\mu_\pi^2\, \bold{\tau}_1 \cdot \bold{\tau}_2 \times
\nonumber \\ & & 
\left\{ \rule{0cm}{\thisheight} \frac{2M_\Lambda}{3M_\Lambda+M}\,   
\left[ f_S(r,\tilde{\mu}_\rho)\, 
\left( 3 K^6_\rho\,  
+ 2 (K^{10}_\rho - K^9_\rho)\, \bold{\sigma}_1\cdot\bold{\sigma}_2 \right)
\right. \right.
\nonumber \\ & & \left. {} \quad 
- (K^{10}_\rho - K^9_\rho)\, f_T(r,\tilde{\mu}_\rho)\, 
S_{12}(\hat{\bold{r}}) \right.
\nonumber \\ & & \left. {} \quad 
- f_V(r,\tilde{\mu}_\rho)\, \left( 
K^{12}_\rho\, \bold{\sigma}_1\times\bold{\sigma}_2 
+ i K^{11}_\rho\, \bold{\sigma}_1 \right) \cdot \hat{\bold{r}} \right]
\nonumber \\ & & \left. {}
- \frac{f_V(r,\tilde{\mu}_\rho)}{r}\, \left[ 
K^6_\rho\, \bold{r}\cdot\bold{\nabla} 
+ K^7_\rho\, \bold{\sigma}_1\cdot\bold{l} 
+ K^8_\rho\, \bold{\sigma}_2\cdot\bold{l}
\right. \right. 
\nonumber \\ & & \left. {} \quad 
+ (K^{10}_\rho - K^9_\rho)\, \bold{\sigma}_1\cdot\bold{\sigma}_2 \; 
\bold{r}\cdot\bold{\nabla}
\right. 
\nonumber \\ & & \left. {} \quad 
+ K^9_\rho\, \bold{\sigma}_2\cdot\bold{r} \; \bold{\sigma}_1\cdot\bold{\nabla} 
- K^{10}_\rho\, \bold{\sigma}_1\cdot\bold{r} \; 
\bold{\sigma}_2\cdot\bold{\nabla}
\right] 
\nonumber \\ & & \left. {} 
+ f_C(r,\tilde{\mu}_\rho)\, \left( 
K^{12}_\rho\, \bold{\sigma}_1\times\bold{\sigma}_2 
+ i K^{11}_\rho\, \bold{\sigma}_1 \right) \cdot \bold{\nabla} 
\rule{0cm}{\thisheight} \right\} \,.
\label{fullV1rhoc} 
\end{eqnarray}

Under approximation \Ref{len}, the only surviving coefficients for the nonlocal 
potential would be 
\begin{eqnarray}
\bar{K}^7_\rho &=& B^V_\rho \, g^V_{NN\rho} \, \frac{1}{4\bar{M}}  
\left( \frac{1}{\bar{M}} + \frac{2}{M} \right) 
+ \frac{B^T_\rho}{2\bar{M}} \, g^V_{NN\rho} \,
\left( \frac{1}{\bar{M}} + \frac{1}{M} \right) \,,
\nonumber \\
\bar{K}^8_\rho &=& B^V_\rho \, g^V_{NN\rho} \, \frac{1}{4M}  
\left( \frac{1}{M} + \frac{2}{\bar{M}} \right) 
\, + \;B^V_\rho \, \frac{g^T_{NN\rho}}{2M} \, 
\left( \frac{1}{M} + \frac{1}{\bar{M}} \right) \,,
\nonumber \\
\bar{K}^{11}_\rho &=&  A_\rho \, g^V_{NN\rho} \,
\left( \frac{1}{M} + \frac{1}{\bar{M}} \right) \,,
\end{eqnarray}
and Eq.~\Ref{fullV1rhoc} would reduce to 
\settoheight{\thisheight}{$ \displaystyle \frac{f_V(r,\mu_\rho)}{r} $}
\begin{eqnarray}
\hat{\bar{V}}^{(1)}_\rho(\bold{r}) &=&  
{} - G_F\,\mu_\pi^2\, \bold{\tau}_1 \cdot \bold{\tau}_2 
\left[ \frac{f_V(r,\mu_\rho)}{r}  
\left( \bar{K}^7_\rho\, \bold{\sigma}_1\cdot\bold{l}  
     + \bar{K}^8_\rho\, \bold{\sigma}_2\cdot\bold{l}  \right) 
     \right. 
\nonumber \\ 
 & & \left. \rule{0cm}{\thisheight} 
 {} - i \bar{K}^{11}_\rho\, f_C(r,\mu_\rho)\, 
\bold{\sigma}_1\cdot\bold{\nabla}
+ \frac{i}{2} \bar{K}^{11}_\rho\, f_V(r,\mu_\rho)\, 
 \bold{\sigma}_1\cdot\hat{\bold{r}} \right]\,.
 \label{avV1rhoc} 
\end{eqnarray}

\subsubsection{Extension to $\eta$ and $\omega$ exchanges} \label{etaomega} 

If one remembers to make the replacement 
$\bold{\tau}_1 \cdot \bold{\tau}_2 \to 1$, the results obtained above for the 
$\pi$ and $\rho$ mesons can be straightforwardly extended to the $\eta$ and 
$\omega$, respectively, and the corresponding expressions need not be reproduced 
here. Let us just mention that the ratios $\tilde{\mu}_\eta/\mu_\eta = 0.985$ 
and $\tilde{\mu}_\omega/\mu_\omega = 0.992$ are very close to unity and, 
consequently, as happened for the $\rho$, the effects of the reduction of the 
effective mass are much less important for these mesons than they are for the 
pion.

\subsection{Contributions of strange mesons}\label{str}

For the strange mesons, the weak and strong vertices in Fig.~\ref{ampc} are 
interchanged with respect to those for the nonstrange ones, \ie,
\begin{eqnarray}
a = W, & & \quad b = S \qquad \mbox{(nonstrange mesons)},
\nonumber \\ 
a = S, & & \quad b = W \qquad \mbox{(strange mesons)}.
\end{eqnarray}

For the kaon, the effective Hamiltonian for the strong coupling is 
\cite[\emph{Eq.(28)}]{Pa97}
\begin{equation} \label{hsk}
\mathcal{H}^S_{\Lambda N K} = i g_{\Lambda N K}\, \bar{\psi}_N \gamma_5\,  
\phi^{(K)}\, \psi_\Lambda ,
\end{equation}
while, for the weak one, it is \cite[\emph{Eq.(29)}]{Pa97}
\begin{eqnarray} 
\mathcal{H}^W_{NNK} &=&  i G_F\, \mu_\pi^2\, \bar{\psi}_N 
\left\{  \left[ C^{PV}_{K} \spurion  
\left(\phi^{(K)}\right)^\dag + D^{PV}_{K}  
\left(\phi^{(K)}\right)^\dag \spurion \right] \right. 
\nonumber \\
& &
\left. {} +  \gamma_{5} 
\left[ C^{PC}_{K} \spurion  \left(\phi^{(K)}\right)^\dag + 
D^{PC}_{K}  
\left(\phi^{(K)}\right)^\dag \spurion \right] 
\right\} \psi_N .
\label{hwk} 
\end{eqnarray}
For the $K^*$, we have, for the strong coupling \cite[\emph{Eq.(38)}]{Pa97},
\begin{equation} \label{hskst}
\mathcal{H}^S_{\Lambda NK^*} =  \bar{\psi}_N 
\left[ \left( g^V_{\Lambda NK^*}\, \gamma^\nu 
+ g^T_{\Lambda NK^*}\, \frac{\sigma^{\mu\nu}\partial_\mu}{2\bar{M}} \right)\, 
\phi^{(K^*)}_{\,\nu} \right]\, \psi_\Lambda , 
\end{equation}
and for the weak one \cite[\emph{Eq.(39)}]{Pa97}, 
\begin{eqnarray} 
\mathcal{H}^W_{NNK^*} &=&  G_F\, \mu_\pi^2\, \bar{\psi}_N 
\left\{ \gamma^\nu \left[ C^{PC,V}_{K^*} \spurion  
\left(\phi^{(K^*)}_{\,\nu}\right)^\dag + D^{PC,V}_{K^*}  
\left(\phi^{(K^*)}_{\,\nu}\right)^\dag \spurion \right] \right. 
\nonumber \\
& & 
{} + \frac{\sigma^{\mu\nu}\partial_\mu}{2M} 
\left[ C^{PC,T}_{K^*} \spurion  \left(\phi^{(K^*)}_{\,\nu}\right)^\dag + 
D^{PC,T}_{K^*}  
\left(\phi^{(K^*)}_{\,\nu}\right)^\dag \spurion \right] 
\nonumber \\
& &
\left. {} + \gamma^\nu \gamma_{5} 
\left[ C^{PV}_{K^*} \spurion  \left(\phi^{(K^*)}_{\,\nu}\right)^\dag + 
D^{PV}_{K^*}  
\left(\phi^{(K^*)}_{\,\nu}\right)^\dag \spurion \right] 
\right\} \psi_N .  
\label{hwkst}
\end{eqnarray}
We again follow the parametrization adopted in Ref.~\cite{Pa97}, and collect the 
numerical values in Table~\ref{strange} for convenience.

%
\settoheight{\thisheight}{ $ \displaystyle D^{PV}_K $ }
\begin{table}[!ht]   
\vspace{.5\baselineskip}
\begin{center}
\caption{\label{strange}
Coupling constants, masses ($\mu_i$) and cutoff parameters ($\Lambda_i$) for the 
strange mesons. The weak couplings are in units of $G_F\,\mu_\pi^2$.  Adapted 
from Ref.~\cite{Pa97}.}
\vspace{.5\baselineskip}

\begin{tabular*}{\textwidth}{@{\extracolsep{\fill}}|clllcc|} 
\hline
\hline 
Meson & \multicolumn{3}{c}{Coupling Constants} & $\mu_i$ & $\Lambda_i$ \\
\cline{2-4}
$i$ & \multicolumn{2}{c}{Weak} & \multicolumn{1}{c}{Strong} & [MeV] & [GeV] \\
\cline{2-3}
 & \multicolumn{1}{c}{$PV$} & \multicolumn{1}{c}{$PC$} & & & \\ 
\hline
& & & & & \\
$K$ &
$C^{PV}_K = 0.76$ & $C^{PC}_K = -18.9$ & $g_{\Lambda NK} = -14.1$ 
& 495.8 & 1.20 \\
& $D^{PV}_K = 2.09$ & $D^{PC}_K = 6.63$ & \rule{0cm}{1.5\thisheight}  
& & \\
& & & & & \\  
$K^*$ &
$C^{PV}_{K^*} = -4.48$ & $C^{PC,V}_{K^*} = -3.61$ & $g^V_{\Lambda NK^*} = -5.47$ 
& 892.4 & 2.20 \\
& & $C^{PC,T}_{K^*} = -17.9$ & $g^T_{\Lambda NK^*} = -11.9$
& \rule{0cm}{1.5\thisheight} & \\
& $D^{PV}_{K^*} = 0.60$ & $D^{PC,V}_{K^*} = -4.89$ & \rule{0cm}{1.5\thisheight}  
& & \\
& \rule{0cm}{1.5\thisheight} & $D^{PC,T}_{K^*} = 9.30$ &   
& & \\
& & & & & \\  
\hline\hline 
\end{tabular*} 
\end{center}
\end{table}

These mesons are isodoublets, and in terms of their different charge states we 
can write  
\[  
\phi^{(K)} \equiv { \phi^{(K^+)} \choose \phi^{(K^0)} } = 
\left[ \phi^{(K^+)}\; \tau_+ + \phi^{(K^0)} \right] \spurion \, , 
\]  
\[  
\left(\phi^{(K)}\right)^\dag \spurion = \left(\phi^{(K^0)}\right)^\dag \, ,
\]  
\begin{equation}
\spurion\left(\phi^{(K)}\right)^\dag = 
\left(\phi^{(K^+)}\right)^\dag\; \tau_- 
+ \frac{1}{2}\left(\phi^{(K^0)}\right)^\dag\; (1 - \tau_0)
\end{equation}
for the kaon, and similar equations for the $K^*$. As a result, when applying 
the Feynman rules to compute the transition potential as explained in 
Subsection~\ref{general}, the isospurion will permit the introduction of isospin 
operators of the form
\begin{equation}
I = \frac{1}{2}\, C
\left( 1 + \bold{\tau}_1 \cdot \bold{\tau}_2 \right) + D 
\end{equation}
for each of the different couplings in Eqs.~\Ref{hwk} and \Ref{hwkst}. 
Explicitly, they are 
\begin{eqnarray}
I^{PV}_{K} &=& \frac{1}{2}\, C^{PV}_{K} 
\left( 1 + \bold{\tau}_1 \cdot \bold{\tau}_2 \right) + D^{PV}_{K} ,
\nonumber \\
I^{PC}_{K} &=& \frac{1}{2}\, C^{PC}_{K} 
\left( 1 + \bold{\tau}_1 \cdot \bold{\tau}_2 \right) + D^{PC}_{K} \, , 
\end{eqnarray}
for the kaon, and 
\begin{eqnarray}
I^{PC,V}_{K^*} &=& \frac{1}{2}\, C^{PC,V}_{K^*} 
\left( 1 + \bold{\tau}_1 \cdot \bold{\tau}_2 \right) + D^{PC,V}_{K^*} ,
\nonumber \\
I^{PC,T}_{K^*} &=& \frac{1}{2}\, C^{PC,T}_{K^*} 
\left( 1 + \bold{\tau}_1 \cdot \bold{\tau}_2 \right) + D^{PC,T}_{K^*} ,
\nonumber \\
I^{PV}_{K^*} &=& \frac{1}{2}\, C^{PV}_{K^*} 
\left( 1 + \bold{\tau}_1 \cdot \bold{\tau}_2 \right) + D^{PV}_{K^*} \, ,
\end{eqnarray}
for the $K^*$.
It then becomes apparent that each such coupling will give a contribution 
proportional to $\frac{1}{2}\,C + D$ to the isoscalar potential and a similar 
one  proportional to $\frac{1}{2}\,C$ to the isovector potential.

\subsubsection{One $K$ exchange} \label{kaon} 

The contribution to the local nonrelativistic transition potential in momentum 
space due to this meson is   
\begin{equation} \label{vk}
V^{(0)}_K(\bold{q}) = \left(1+\frac{M_0}{M_q}\right)^{-1} 
G_F\,\mu_\pi^2 \,
\frac{g_{\Lambda NK}}{2\check{M}} 
\left( I_K^{PV} - \frac{I_K^{PC}}{2M}\, 
\bold{\sigma}_2 \cdot \bold{q} \right) \; 
\frac{\bold{\sigma}_1 \cdot \bold{q}}{\bold{q}^2 + \tilde{\mu}_K^2} \, .
\end{equation}
Comparing this with the result that would have been obtained under 
approximation \Ref{len}, namely%
\footnote{This differs from Eqs. (24) and (31) of Ref.~\cite{Pa97} in the sign 
of $I_K^{PV}$ and the interchange of $\bold{\sigma}_1$ and $\bold{\sigma}_2$.},  
\begin{equation}
\bar{V}^{(0)}_K(\bold{q}) = G_F\,\mu_\pi^2 \,
\frac{g_{\Lambda NK}}{2\bar{M}} 
\left( I_K^{PV} - \frac{I_K^{PC}}{2M}\, 
\bold{\sigma}_2 \cdot \bold{q} \right) \; 
\frac{\bold{\sigma}_1 \cdot \bold{q}}{\bold{q}^2 + \mu_K^2} \,,
\end{equation}
and noticing that the two correction factors in Eq.~\Ref{cf12} as well as  
\[
\tilde{\mu}_{K}/\mu_{K} = 0.982
\]
are very close to unity, one can see that, for the kaon, only very 
small corrections will result in the local contributions from the more accurate 
treatment of the kinematics. 
The expression for the potential \Ref{vk} in coordinate space is 
\settoheight{\thisheight}{ $ \displaystyle \frac{I^{PC}_{K}}{2M} $ }
\begin{eqnarray}
V^{(0)}_K(\bold{r}) &=&  \left(1+\frac{M_0}{M_q}\right)^{-1} 
G_F\,\mu_\pi^2 \;
\frac{g_{\Lambda NK}}{2\check{M}}  
\left[ \rule{0cm}{\thisheight}   
i I^{PV}_{K}\, f_V(r,\tilde{\mu}_K)\, \bold{\sigma}_1 \cdot \hat{\bold{r}}
\right.
\nonumber \\ & & \left. {}
+ \frac{I^{PC}_{K}}{2M}\, f_S(r,\tilde{\mu}_K)\, 
\bold{\sigma}_1 \cdot \bold{\sigma}_2  
+ \frac{I^{PC}_{K}}{2M}\, f_T(r,\tilde{\mu}_K)\, S_{12}(\hat{\bold{r}})  
\right] \,, 
\end{eqnarray}
and, under approximation \Ref{len}, this becomes
\begin{eqnarray}
\bar{V}^{(0)}_K(\bold{r}) &=&  G_F\,\mu_\pi^2 \;
\frac{g_{\Lambda NK}}{2\bar{M}}  
\left[ \rule{0cm}{\thisheight}   
i I^{PV}_{K}\, f_V(r,\mu_K)\, \bold{\sigma}_1 \cdot \hat{\bold{r}} 
\right.
\nonumber \\ & & \left. {}
+ \frac{I^{PC}_{K}}{2M}\, f_S(r,\mu_K)\, \bold{\sigma}_1 \cdot \bold{\sigma}_2\;
+ \frac{I^{PC}_{K}}{2M}\, f_T(r,\mu_K)\, S_{12}(\hat{\bold{r}}) 
\right] . 
\end{eqnarray}
%


Starting with the first-order nonlocality coefficient in momentum space in 
Eq.~\Ref{Vmom}, we have, for the kaon, 
\begin{equation} \label{V1K}
\bold{V}^{(1)}_K(\bold{q}) = \left(1+\frac{M_0}{M_q}\right)^{-1} 
G_F\,\mu_\pi^2 \,
\frac{g_{\Lambda NK}}{2\grave{M}} \left( I_K^{PV} -  \frac{I_K^{PC}}{2M}\, 
\bold{\sigma}_2 \cdot \bold{q}\right) \; 
\frac{\bold{\sigma}_1}{\bold{q}^2 + \tilde{\mu}_K^2} \,, 
\end{equation}
which, in the coordinate representation, becomes 
\begin{eqnarray}
\bold{V}^{(1)}_K(\bold{r}) &=& \left( 1 + \frac{M_0}{M_q} \right)^{-1} \; 
G_F\,\mu_\pi^2 \;\frac{g_{\Lambda NK}}{2\grave{M}}  \; 
\left[ \rule{0cm}{\thisheight} I_K^{PV}\, f_C(r,\tilde{\mu}_K)\, 
\bold{\sigma}_1 \right.
\nonumber\\
& & \left. {}  
- i \frac{I_K^{PC}}{2M}\, f_V(r,\tilde{\mu}_K)\,
(\bold{\sigma}_2 \cdot \hat{\bold{r}})\bold{\sigma}_1 \, \right] \,.
\end{eqnarray}
Introducing this into Eq.~\Ref{Vnl} yields, for the first-order nonlocal 
potential,  
\begin{eqnarray}
\hat{V}^{(1)}_K(\bold{r}) &=&  \left( 1 + \frac{M_0}{M_q} \right)^{-1} \; 
G_F\,\mu_\pi^2 \;
\frac{g_{\Lambda NK}}{2\grave{M}} \; \times
\nonumber \\  & & {}  
\left\{ 
\frac{2M_\Lambda}{3M_\Lambda+M} \left[ \frac{I_K^{PC}}{2M}\, 
f_S(r,\tilde{\mu}_K )\, \bold{\sigma}_1\cdot\bold{\sigma}_2 \;
+ \, \frac{I_K^{PC}}{2M}\,f_T(r,\tilde{\mu}_K )\;S_{12}(\hat{\bold{r}}) 
\right.\right.
\nonumber\\ & & \left.
\qquad {}  
+ i I_K^{PV}\,f_V(r,\tilde{\mu}_K )\; (\bold{\sigma}_1\cdot\hat{\bold{r}}) 
\rule{0cm}{\thisheight} \right]
\nonumber \\ & & \left. {} 
- i I_K^{PV}\, f_C(r,\tilde{\mu}_K )\, (\bold{\sigma}_1\cdot{\bold{\nabla}})
- \frac{I_K^{PC}}{2M}\, f_V(r,\tilde{\mu}_K )\, 
(\bold{\sigma}_2\cdot\hat{\bold{r}}) (\bold{\sigma}_1\cdot{\bold{\nabla}})  
\right\} \,.
\label{fullV1kaonc}
\end{eqnarray}
As already stated, the mass averaging approximation \Ref{len} would set 
$1/\grave{M}$, defined in Eq.~\Ref{graveM}, to zero. Therefore,  
there would be no first-order nonlocality correction for the kaon under this 
approximation, \ie, 
\begin{equation} \label{avV1kaonc}
\hat{\bar{V}}^{(1)}_K(\bold{r}) = 0 \,.
\end{equation}

\subsubsection{One $K^*$ exchange} \label{kstar}  

For one-$K^*$-exchange, the local nonrelativistic transition potential in 
momentum space is 
\begin{eqnarray}  
V^{(0)}_{K^*}(\bold{q}) &=& \left(1+\frac{M_0}{M_q}\right)^{-1} 
G_F\,\mu_\pi^2\; 
\Bigm[ \hat{K}^1_{K^*} - \hat{K}^2_{K^*}\, \bold{q}^2 
\nonumber \\
& & {} - \hat{K}^3_{K^*}\, (\bold{\sigma}_1 \times \bold{q}) \cdot 
        (\bold{\sigma}_2 \times \bold{q})   
\nonumber \\
& & {} - i\,\hat{K}^4_{K^*}\, (\bold{\sigma}_1 \times \bold{\sigma}_2) \cdot 
\bold{q}   
 + \hat{K}^5_{K^*} \, (\bold{\sigma}_2 \cdot \bold{q}) \Bigm]  
\frac{1}{\bold{q}^2 + \tilde{\mu}_{K^*}} \,, 
\label{vkst}              
\end{eqnarray}
where we have introduced the isospin operators
\begin{eqnarray}
\hat{K}^1_{K^*} &=& g^V_{\Lambda N K^*} \,I_{K^*}^{PC,V}\, ,
\nonumber \\
\hat{K}^2_{K^*} &=& g^V_{\Lambda N K^*} \,I_{K^*}^{PC,V}\,  
                 \left[ \left( \frac{1}{4M} \right)^2 
                      + \left( \frac{1}{4\check{M}} \right)^2  
                 + \frac{1}{2} \left( \frac{1}{M_q} \right)^2  \right] 
\nonumber \\       
& & {} + \left( \frac{g^T_{\Lambda N K^*}}{2\bar{M}}\,                           
         \frac{I_{K^*}^{PC,V}}{2\check{M}}
       + \frac{g^V_{\Lambda N K^*}}{2M}\, \frac{I_{K^*}^{PC,T}}{2M}\, \right)\,
         \left( 1 + \frac{M_0}{M_q} \right)
\nonumber \\          
& & {} + \frac{g^T_{\Lambda N K^*}}{2\bar{M}}\, \frac{I_{K^*}^{PC,T}}{2M} \, 
         \left( \frac{M_0^2}{4M\check{M}} \right) ,
\nonumber \\
\hat{K}^3_{K^*} &=& \left[ \frac{I_{K^*}^{PC,V}}{2M} +  
\frac{I_{K^*}^{PC,T}}{2M}\,\left( 1 + \frac{M_0}{M_q} \right) \right]   
\left[ \frac{g^V_{\Lambda N K^*}}{2\check{M}} + \frac{g^T_{\Lambda N 
K^*}}{2\bar{M}}\,
\left( 1 + \frac{M_0}{M_q} \right) \right] ,
\nonumber \\
\hat{K}^4_{K^*} &=&  I_{K^*}^{PV} \,
\left[ \frac{g^V_{\Lambda N K^*}}{2\check{M}} + \frac{g^T_{\Lambda N 
K^*}}{2\bar{M}}
\left( 1 + \frac{M_0}{M_q} \right) \right] ,
\nonumber \\
\hat{K}^5_{K^*} &=& I_{K^*}^{PV} \, \frac{g^T_{\Lambda N K^*}}{2\bar{M}} 
\left(\frac{M_0}{2\check{M}}\right) .
\end{eqnarray}

The corresponding potential under approximation \Ref{len}, 
$\bar{V}^{(0)}_{K^*}(\bold{q})$, can be obtained from Eq.~\Ref{vkst} through 
the substitutions: 
\begin{equation} \label{ksubs}
V^{(0)}_{K^*} \rightarrow \bar{V}^{(0)}_{K^*} \,,
\qquad \hat{K}^j_{K^*} \rightarrow \lhat{\bar{K}^j_{K^*}} \;\; 
(j=1 \mbox{--} 5)\,,
\qquad M_0/M_q \rightarrow 0\,, 
\qquad \tilde{\mu}_{K^*} \rightarrow \mu_{K^*}\,,
\end{equation}
with
\begin{eqnarray}
\lhat{\bar{K}^1_{K^*}} &=& g^V_{\Lambda N K^*} \, I_{K^*}^{PC,V} \,,
\nonumber \\
\lhat{\bar{K}^2_{K^*}} &=& g^V_{\Lambda N K^*} \, I_{K^*}^{PC,V} \,  
\left[ \left( \frac{1}{4M} \right)^2 +
       \left( \frac{1}{4\bar{M}} \right)^2  \right] + 
       \frac{g^T_{\Lambda N K^*}}{2\bar{M}}\,\frac{I_{K^*}^{PC,V}}{2\bar{M}} +
       \frac{g^V_{\Lambda N K^*}}{2M}\, \frac{I_{K^*}^{PC,T}}{2M} \,,
\nonumber \\
\lhat{\bar{K}^3_{K^*}} &=& \left(\frac{I_{K^*}^{PC,V} + 
I_{K^*}^{PC,T}}{2M}\right) \,
            \left(\frac{g^V_{\Lambda N K^*} +                         
g^T_{\Lambda N K^*}}{2\bar{M}}\right) \,,
\nonumber \\
\lhat{\bar{K}^4_{K^*}} &=& I_{K^*}^{PV} \, 
              \left(\frac{g^V_{\Lambda N K^*} +                       
g^T_{\Lambda N K^*}}{2\bar{M}}\right) \,, 
\nonumber \\
\lhat{\bar{K}^5_{K^*}} &=& 0 \,.
\end{eqnarray}

By an analysis very similar to the one performed for the $\rho$ meson 
and noticing that 
\[
\tilde{\mu}_{K^*}/\mu_{K^*} = 0.994
\]
is also very close to unity, 
one concludes that, again in the present case, only very small corrections will 
result in the local contributions from the more accurate treatment of the 
kinematics.

For completeness, we give below the expression for the potential \Ref{vkst} in 
coordinate space:
\begin{eqnarray} 
V^{(0)}_{K^*}(\bold{r}) &=& \left(1+\frac{M_0}{M_q}\right)^{-1} 
G_F\,\mu_\pi^2\; 
\left\{ 
\hat{K}^1_{K^*}  f_C(r,\tilde{\mu}_{K^*})
+ 3 \hat{K}^2_{K^*} f_S(r,\tilde{\mu}_{K^*}) 
\right.
\nonumber \\
& & {} + 2 \hat{K}^3_{K^*}\, f_S(r,\tilde{\mu}_{K^*})\, 
\bold{\sigma}_1 \cdot \bold{\sigma}_2  
- \hat{K}^3_{K^*}\, f_T(r,\tilde{\mu}_{K^*})\, S_{12}(\hat{\bold{r}}) 
\nonumber \\ 
& & \left. {} + f_V(r,\tilde{\mu}_{K^*}) 
\left[ \hat{K}^4_{K^*}\, (\bold{\sigma}_1 \times \bold{\sigma}_2)  
+ i \hat{K}^5_{K^*}\, \bold{\sigma}_2 \right] \cdot \hat{\bold{r}} 
\right\} .  
\label{vkstc}
\end{eqnarray}
Once more, the corresponding potential under approximation \Ref{len} can be 
obtained from Eq.~\Ref{vkstc} by means of the substitutions \Ref{ksubs}.

The first-order nonlocality coefficient in momentum space,   
appearing in Eq.~\Ref{Vmom},  
for this meson, is 
\begin{eqnarray} 
\bold{V}^{(1)}_{K^*}(\bold{q}) &=& - \left(1+\frac{M_0}{M_q}\right)^{-1}
G_F\,\mu_\pi^2\, \times 
\nonumber \\ & & 
\left[ {\hat{K}}^6_{K^*}\, \bold{q}
      - i {\hat{K}}^7_{K^*}\, \bold{\sigma}_1 \times \bold{q}   
      - i {\hat{K}}^8_{K^*}\, \bold{\sigma}_2 \times \bold{q}  
\right.
\nonumber \\ & & \left. {} 
- {\hat{K}}^{9}_{K^*}\, (\bold{\sigma}_1\times\bold{q}) \times \bold{\sigma}_2 
+ {\hat{K}}^{10}_{K^*}\, (\bold{\sigma}_2\times\bold{q}) \times \bold{\sigma}_1
\right.   
\nonumber \\ & & \left. {}     
      - {\hat{K}}^{11}_{K^*}\, \bold{\sigma}_2
    + i {\hat{K}}^{12}_{K^*}\, \bold{\sigma}_1\times\bold{\sigma}_2  \right] 
\frac{1}{\bold{q}^2 + \tilde{\mu}_{K^*}^2} \,,
\label{Vks1m}
\end{eqnarray}
where 
\begin{eqnarray}
{\hat{K}}^6_{K^*} &=& g^V_{\Lambda NK^*}\, I^{PC,V}_{K^*}\,  
\frac{1}{16\grave{M}} \, 
\left( \frac{3}{\check{M}} - \frac{1}{M} \right) 
+ g^V_{\Lambda NK^*}\, \frac{I^{PC,T}_{K^*}}{2M}\, \frac{M_0}{4M\acute{M}}  
\nonumber \\ & & {} 
+ \frac{g^T_{\Lambda NK^*}}{2\bar{M}}\, I^{PC,V}_{K^*}\, 
\left[ \frac{1}{2\grave{M}}\, \left(1+\frac{M_0}{M_q}\right) - 
\frac{M_0}{2M\check{M}} \right]
+ \frac{g^T_{\Lambda NK^*}}{2\bar{M}}\, \frac{I^{PC,T}_{K^*}}{2M}\, 
\frac{M_0^2}{4M\grave{M}} \,,
\nonumber \\
{\hat{K}}^7_{K^*} &=& g^V_{\Lambda NK^*}\, I^{PC,V}_{K^*}\, 
\left[ \frac{1}{8\check{M}}\, \left( \frac{2}{M} + \frac{1}{\acute{M}} \right) 
+  \frac{1}{8M\acute{M}} \right] 
+ g^V_{\Lambda NK^*}\, \frac{I^{PC,T}_{K^*}}{2M}\, \frac{M_0}{4M\grave{M}}  
\nonumber \\ & & {} 
+ \frac{g^T_{\Lambda NK^*}}{2\bar{M}}\, I^{PC,V}_{K^*}\, 
\frac{1}{2} \left( \frac{2}{M} + \frac{1}{\acute{M}} \right) 
\left(1+\frac{M_0}{M_q}\right) 
+ \frac{g^T_{\Lambda NK^*}}{2\bar{M}}\, \frac{I^{PC,T}_{K^*}}{2M}\, 
\frac{M_0^2}{4M\acute{M}} \,,
\nonumber \\
{\hat{K}}^8_{K^*} &=& g^V_{\Lambda NK^*} I^{PC,V}_{K^*}  
\frac{1}{4M} \left( \frac{1}{M} + \frac{1}{\acute{M}} \right) 
+ g^V_{\Lambda NK^*} \frac{I^{PC,T}_{K^*}}{2M}\,   
\frac{1}{2} \left( \frac{2}{M} + \frac{1}{\acute{M}} \right) 
\left(1+\frac{M_0}{M_q}\right) 
\nonumber \\ & & {} 
+ \frac{g^T_{\Lambda NK^*}}{2\bar{M}}\, I^{PC,V}_{K^*}\, \frac{M_0}{4M\grave{M}}
+ \frac{g^T_{\Lambda NK^*}}{2\bar{M}}\, \frac{I^{PC,T}_{K^*}}{2M}\, 
\frac{M_0}{2}\,
\left[ \frac{1}{\grave{M}}\, \left(1+\frac{M_0}{M_q}\right) 
+ \frac{M_0}{M\check{M}} \right] \,,
\nonumber \\
{\hat{K}}^{9}_{K^*} &=&  g^V_{\Lambda NK^*}\, \frac{I^{PC,T}_{K^*}}{2M}\, 
\frac{M_0}{2M\check{M}}
+ \frac{g^T_{\Lambda NK^*}}{2\bar{M}}\, \frac{I^{PC,T}_{K^*}}{2M}\, 
\frac{M_0}{M}\,
\left(1+\frac{M_0}{M_q}\right) \,,
\nonumber \\
{\hat{K}}^{10}_{K^*} &=& g^V_{\Lambda NK^*}\, I^{PC,V}_{K^*}\, 
\frac{1}{4M\grave{M}}\,  
+ g^V_{\Lambda NK^*}\, \frac{I^{PC,T}_{K^*}}{2M}\, 
\frac{1}{2\grave{M}}\, \left(1+\frac{M_0}{M_q}\right) 
\nonumber \\ & & {} 
+ \frac{g^T_{\Lambda NK^*}}{2\bar{M}}\, I^{PC,V}_{K^*}\, \frac{M_0}{4M\acute{M}}
+ \frac{g^T_{\Lambda NK^*}}{2\bar{M}}\, \frac{I^{PC,T}_{K^*}}{2M}\, 
\frac{M_0}{2\acute{M}}\, \left(1+\frac{M_0}{M_q}\right) \,,
\nonumber \\
{\hat{K}}^{11}_{K^*} &=&  g^V_{\Lambda NK^*}\, I^{PV}_{K^*}\, 
\frac{1}{2} \left( \frac{2}{M} + \frac{1}{\acute{M}} \right) 
+ \frac{g^T_{\Lambda NK^*}}{2\bar{M}} \, I^{PV}_{K^*} \, 
\frac{M_0}{2\grave{M}} \,,
\nonumber \\
{\hat{K}}^{12}_{K^*} &=& g^V_{\Lambda NK^*}\, I^{PV}_{K^*}\, 
\frac{1}{2\grave{M}} + \frac{g^T_{\Lambda NK^*}}{2\bar{M}} \, I^{PV}_{K^*} \, 
\frac{M_0}{2\acute{M}} \,,
\end{eqnarray}
with $1/\grave{M}$ and $1/\acute{M}$ as defined in Eqs. \Ref{graveM} and 
\Ref{acuteM}.
To get the first-order nonlocality correction  $\hat{V}^{(1)}_{K^*}(\bold{r})$, 
we need first to change \Ref{Vks1m} to the coordinate representation. This gives
%
\settoheight{\thisheight}{$ \displaystyle \frac{f_V(r,\mu_\rho)}{r} $}
\begin{eqnarray} 
\bold{V}^{(1)}_{K^*}(\bold{r}) &=& - \left(1+\frac{M_0}{M_q}\right)^{-1}
G_F\,\mu_\pi^2\, \times 
\nonumber \\ & & 
\left\{ \frac{f_V(r,\tilde{\mu}_{K^*})}{r}\,
\left[ i {\hat{K}}^6_{K^*}\, \bold{r} 
       + {\hat{K}}^7_{K^*}\, \bold{\sigma}_1 \times \bold{r} 
       + {\hat{K}}^8_{K^*}\, \bold{\sigma}_2 \times \bold{r}  
\right. \right.
\nonumber \\ & & \left. {}
   - i {\hat{K}}^{9}_{K^*}\, (\bold{\sigma}_1\times\bold{r}) \times 
\bold{\sigma}_2 
   + i {\hat{K}}^{10}_{K^*}\, (\bold{\sigma}_2\times\bold{r}) \times 
\bold{\sigma}_1 
\right]   
\nonumber \\ & & \left. {} 
- f_C(r,\tilde{\mu}_{K^*})\, \left[     
        {\hat{K}}^{11}_{K^*}\, \bold{\sigma}_2
    - i {\hat{K}}^{12}_{K^*}\, \bold{\sigma}_1\times\bold{\sigma}_2 \right] 
\rule{0cm}{\thisheight} \right\} \,. 
\label{Vks1c}
\end{eqnarray}
Introducing \Ref{Vks1c} into Eq.~\Ref{Vnl}, we obtain, finally,
\begin{eqnarray} 
\hat{V}^{(1)}_{K^*}(\bold{r}) &=&  \left(1+\frac{M_0}{M_q}\right)^{-1}
G_F\,\mu_\pi^2\, \times
\nonumber \\ & & 
\left\{ \rule{0cm}{\thisheight} \frac{2M_\Lambda}{3M_\Lambda+M}\,   
\left[ f_S(r,\tilde{\mu}_{K^*})\, 
\left( 3 {\hat{K}}^6_{K^*}\,  
+ 2 ({\hat{K}}^{10}_{K^*} - {\hat{K}}^{9}_{K^*})\, 
\bold{\sigma}_1\cdot\bold{\sigma}_2 \right)
\right. \right.
\nonumber \\ & & \left. {} \quad 
- ({\hat{K}}^{10}_{K^*} - {\hat{K}}^{9}_{K^*})\, f_T(r,\tilde{\mu}_{K^*})\, 
S_{12}(\hat{\bold{r}}) \right.
\nonumber \\ & & \left. {} \quad 
+ f_V(r,\tilde{\mu}_{K^*})\, \left( 
{\hat{K}}^{12}_{K^*}\, \bold{\sigma}_1\times\bold{\sigma}_2 
+ i {\hat{K}}^{11}_{K^*}\, \bold{\sigma}_2 \right) \cdot \hat{\bold{r}} \right]
\nonumber \\ & & \left. {}
- \frac{f_V(r,\tilde{\mu}_{K^*})}{r}\, \left[ 
{\hat{K}}^6_{K^*}\, \bold{r}\cdot\bold{\nabla}
+ {\hat{K}}^7_{K^*}\, \bold{\sigma}_1\cdot\bold{l}
+ {\hat{K}}^8_{K^*}\, \bold{\sigma}_2\cdot\bold{l} 
\right. \right. 
\nonumber \\ & & \left. {} \quad 
+ ({\hat{K}}^{10}_{K^*} - {\hat{K}}^{9}_{K^*})\, 
\bold{\sigma}_1\cdot\bold{\sigma}_2 \; 
\bold{r}\cdot\bold{\nabla}
\right. 
\nonumber \\ & & \left. {} \quad 
+ {\hat{K}}^{9}_{K^*}\, \bold{\sigma}_2\cdot\bold{r} \; 
\bold{\sigma}_1\cdot\bold{\nabla}
- {\hat{K}}^{10}_{K^*}\, \bold{\sigma}_1\cdot\bold{r} \; 
\bold{\sigma}_2\cdot\bold{\nabla} 
\right] 
\nonumber \\ & & \left. {} 
- f_C(r,\tilde{\mu}_{K^*})\, \left( 
{\hat{K}}^{12}_{K^*}\, \bold{\sigma}_1\times\bold{\sigma}_2 
+ i {\hat{K}}^{11}_{K^*}\, \bold{\sigma}_2 \right) \cdot \bold{\nabla} 
\rule{0cm}{\thisheight} \right\} \,.
\label{fullV1kstarc} 
\end{eqnarray}

If one assumed that the averaged-mass approximation \Ref{len} could be made, 
several terms in the nonlocal potential would disappear. The only remaining 
coefficients would be 
\begin{eqnarray}
\lhat{\bar{K}^7_{K^*}} &=& g^V_{\Lambda N K^*} \, I_{K^*}^{PC,V} \, 
\frac{1}{4\bar{M}}  
\left( \frac{1}{\bar{M}} + \frac{2}{M} \right) 
\, + \;\frac{g^T_{\Lambda N K^*}}{2\bar{M}} \,I_{K^*}^{PC,V} \, 
\left( \frac{1}{M} + \frac{1}{\bar{M}} \right) \,,
\nonumber \\
\lhat{\bar{K}^8_{K^*}} &=& g^V_{\Lambda N K^*} \, I_{K^*}^{PC,V} \,  
\frac{1}{4M}  
\left( \frac{1}{M} + \frac{2}{\bar{M}} \right) 
+ g^V_{\Lambda N K^*} \,\frac{I_{K^*}^{PC,T}}{2M} \, 
\left( \frac{1}{\bar{M}} + \frac{1}{M} \right) \,,
\nonumber \\
\lhat{\bar{K}^{11}_{K^*}} &=&  \, g^V_{\Lambda N K^*}\,I_{K^*}^{PV}\,
\left( \frac{1}{M} + \frac{1}{\bar{M}} \right) \,,
\end{eqnarray}
and the first order nonlocality correction would reduce to 
\settoheight{\thisheight}{ $ \displaystyle \frac{f_V(r,\mu_{K^*})}{r} $ }
\begin{eqnarray}
\hat{\bar{V}}^{(1)}_{K^*}(\bold{r}) &=&
{} - G_F\,\mu_\pi^2\,  
\left[ \frac{f_V(r,\mu_{K^*})}{r}  
\left( \lhat{\bar{K}^7_{K^*}}\, \bold{\sigma}_1\cdot\bold{l}
     + \lhat{\bar{K}^8_{K^*}}\, \bold{\sigma}_2\cdot\bold{l}   \right) 
     \right. 
\nonumber \\ 
 & & 
 \left. \rule{0cm}{\thisheight} 
{} + i \lhat{\bar{K}^{11}_{K^*}}\, f_C(r,\mu_{K^*})\, 
\bold{\sigma}_2\cdot\bold{\nabla}
-\frac{i}{2} \lhat{\bar{K}^{11}_{K^*}}\, 
f_V(r,\mu_{K^*})\, 
\bold{\sigma}_2 \cdot \hat{\bold{r}} \right]\,.
\label{avV1kstarc} 
\end{eqnarray}

It is interesting to point out that, whereas the first-order nolocal terms are 
systematically omitted in the literature on nonmesonic decay, they are routinely 
included in the closely related domain of strangeness-conserving,   
parity-violating nuclear forces. (See, for instance, Eq. (115) in 
Ref.~\cite{De80}.)
We note, however, that the terms proportional to $\bar{K}^{11}$ and 
$\,\lhat{\!\bar{K}^{11}}$, for the vector mesons, have been recently discussed 
in the literature \cite{Ba02}. 

\subsection{Finite size effects} \label{fse} 

Before closing this section, let us mention a refinement that should always be 
added to the strict OME description we have been developing up to now, 
especially when large momentum transfers are involved, as is the case for 
nonmesonic hypernuclear decays. 
This is the effect of the finite size (FS) of the interacting baryons and mesons 
at each vertex. 

Taking a clue from the OME models for the $NN$ force \cite{Ma87,Ma89}, the FS 
effects are phenomenologically implemented in momentum space by inserting, at 
each vertex in Fig.~\ref{ampc}, a form factor, which we choose to be of the 
monopole type,  
\begin{equation}\label{form}
\frac{\Lambda_i^2 - \tilde{\mu}_i^2}{\bold{q}^2 + \Lambda_i^2} \,,
\end{equation}
where $i$ refers to the 
meson involved and $\Lambda_i$ are the cutoff parameters in Tables 
\ref{nonstrange} and \ref{strange}. This corresponds in coordinate space to 
replacing, in the expressions for the transition potential discussed in 
Subsections \ref{nstr} and \ref{str}, each of the shape functions 
\Ref{fCVST} as follows: 
\begin{equation}\label{shape}
f_N(r,\tilde{\mu}_i) \to f_N(r,\tilde{\mu}_i) - f_N(r,\Lambda_i) + 
\frac{\Lambda_i^2 - \tilde{\mu}_i^2}{2\Lambda_i}\, \frac{\partial}{\partial 
\Lambda_i} 
f_N(r,\Lambda_i) ,
\end{equation}
where $N= C, V, S, T$. When the kinematical effects are ignored, Eqs. \Ref{form} 
and \Ref{shape} should be modified by making $\tilde{\mu}_i \to \mu_i$, thus 
leading to agreement with Ref.~\cite{Pa97}.

In what follows, it is to be understood that these FS effects are always 
included.

\section{Numerical results and discussion}\label{num} 

\subsection{Decay rates} \label{rates}

We present here the numerical results for the different contributions
to the nonmesonic weak decay rates of $^{12}_{\,\;\Lambda}$C. 
We consider, separately, the 
neutron-induced ($n$) and the proton-induced ($p$)  contributions, as well as 
those coming from the parity-conserving ($PC$) and parity-violating ($PV$) 
transitions. All quantities are in units of the free $\Lambda$ decay constant, 
$\Gamma_0 = 2.50 \times 10^{-6}$ eV. The main observables are the total 
nonmesonic decay constant $\Gamma_{nm} = \Gamma_n + \Gamma_p$ and the ratio 
$\Gamma_n/\Gamma_p$, whose experimental estimates are in the ranges 0.89~--~1.14 
and 0.52~--~1.87, respectively, with large error bars \cite{Mo74}--\cite{Ha02}.  
Most, if not all, calculations in the context of OME models give reasonable 
results for $\Gamma_{nm}$ but fail completely for $\Gamma_n/\Gamma_p$. 
However, our main objective here is not 
so much to try to reproduce the experimental values for these observables, but 
rather to assess the relative importance of the kinematical and nonlocality 
effects, usually ignored, in their theoretical prediction.
For simplicity, we restrict the discussion of the nonlocality corrections 
to those of first-order.

For the explicit evaluation of the transition rates, we follow  the approach of 
Ref.~\cite{Ba02}. 
The initial and final nuclear states in Eq.~\Ref{golden} are described in the 
extreme particle-hole model (EPHM), taking as vacuum the simplest possible 
shell-model approximation for the  ground state of $^{12}$C, namely, $1s_{1/2}$ 
and $1p_{3/2}$ orbitals completely filled with neutrons and protons.\footnote{As 
shown in that reference, further sophistication of the nuclear structure 
description has little effect on the nonmesonic decay rates.}  
The $\Lambda$ single-particle state has quantum numbers $j_1=1s_{1/2}$ and the 
nucleon inducing the transition occupies a $j_2 = 1s_{1/2} \mbox{~or~} 1p_{3/2}$ 
orbital. 
Therefore, \cite[\emph{Eqs.(4.2,3)}]{Ba02} 
\begin{equation}\label{states}
\ket{I} = \ket{(j_1 \Lambda)\, (j n)^{-1}; J_I} \,, 
\qquad \qquad 
\ket{F} = \ket{(j n)^{-1}\, (j_2 N)^{-1}; J_F} \,,
\end{equation}
where $J_I=1$, $j=1p_{3/2}$, $N=p\mbox{~or~}n$ for proton- or neutron- induced 
transitions, respectively, and $J_F$ takes all the values allowed by angular 
momentum coupling and (when relevant) antisymmetrization.
Then, changing the momentum variables in Eq.~\Ref{golden} to relative 
($\bold{p}'$) and center-of-mass ($\bold{P}'$) momenta, making a multipole 
decomposition of the corresponding free waves and performing the angular 
integrations, one gets, for $N$-induced transitions, 
\cite[\emph{Eqs.(2.4,9)}]{Ba02} 
\begin{eqnarray}
\lefteqn{
\Gamma_N = \frac{16M^3}{\pi} \sum_{j_2 J_F} 
\int_0^{\Delta_{j_2 N}} d\epsilon'\, 
\sqrt{\epsilon'(\Delta_{j_2 N} - \epsilon')} \;\times
} 
\nonumber \\  & & 
\sum_{l^\prime L'\lambda'S' \atop J'T'M'_T} 
\left|\bra{p'l'P'L'\lambda'S'J'T'M'_T\,, (j n)^{-1}\, (j_2 N)^{-1}; J_F;J_I}
\hat{V}\ket{(j_1 \Lambda)\, (j n)^{-1}; J_I}\right|^2 \,,  \qquad {}
\label{golden1}
\end{eqnarray}
where $T'$ is the total isospin of the two emitted nucleons and the angular 
momentum couplings $\bold{l}'+\bold{L}'=\bold{\lambda}'$, 
$\bold{\lambda}'+\bold{S}'=\bold{J}'$ and $\bold{J}'+\bold{J}_F=\bold{J}_I$ 
are carried out.
One also has $P'=2\sqrt{M\epsilon'}$, $p'=\sqrt{M(\Delta_{j_2 N}-\epsilon')}$ 
and $\Delta_{j_2 N} = M_\Lambda - M + \varepsilon_{j_1\Lambda} +  
\varepsilon_{j_2N}$, where the single-particle energies are taken from 
experiment, according to Table 3 of Ref.~\cite{Ra92}.

After some standard manipulations, Eq.~\Ref{golden1} takes the form 
\cite[\emph{Eqs.(2.13),(4.4)}]{Ba02}
\begin{eqnarray}
\Gamma_N &=& \frac{16M^3}{\pi} \sum_{j_2} 
\int_0^{\Delta_{j_2 N}} d\epsilon'\, 
\sqrt{\epsilon'(\Delta_{j_2 N} - \epsilon')} 
\;\times
\nonumber \\  & & 
\sum_{J'} F^{j_2 N}_{J'} \sum_{l'L'\lambda'S'T'} 
\left| \mathcal{M}(p'l'P'L'\lambda'S'J'T'\,;\, j_1 \Lambda \; j_2 N) \right|^2 
\,,  
\label{golden2}
\end{eqnarray} 
which allows a nice separation between the nuclear structure aspects and those 
of the decay dynamics proper. 
The nuclear structure factor is, in second-quantized notation,  
\begin{equation}\label{structure}
F^{j_2 N}_{J'} = \frac{1}{2J_I+1} \sum_{J_F} 
\left| \Bra{I} 
\left( a^\dagger_{j_2 N}\, a^\dagger_{j_1 \Lambda} \right)_{J'} 
\Ket{F} \right|^2 \,, 
\end{equation}
and its nonzero values, for the nuclear states in Eqs.~\Ref{states}, are: 
$F^{1s_{1/2}\,n}_{0} = F^{1s_{1/2}\,p}_{0} = 1/2$, $F^{1s_{1/2}\,n}_{1} = 
F^{1s_{1/2}\,p}_{1} = 3/2$, $F^{1p_{3/2}\,n}_{1} = 7/4$, $F^{1p_{3/2}\,n}_{2} = 
5/4$, $F^{1p_{3/2}\,p}_{1} = 3/2$ and $F^{1p_{3/2}\,p}_{2} = 5/2$. 
(See Table I in Ref.~\cite{Kr02}.) 
On the other hand, the nuclear matrix element governing the decay is  
\begin{eqnarray}
\lefteqn{
\mathcal{M}(p'l'P'L'\lambda'S'J'T'\,;\, j_1 \Lambda \; j_2 N) = 
} 
\nonumber \\  & & 
\frac{1}{\sqrt{2}} \left[ 1 - (-)^{l'+S'+T'} \right] \, 
\roundbra{p'l'P'L'\lambda'S'J'T'M_T} \hat{V} 
\roundket{j_1 \Lambda \; j_2 N\, J'} \,,
\label{matrix}
\end{eqnarray}
where $\roundbra{\cdots}\hat{V}\roundket{\cdots}$ is a direct matrix element 
and the factor in front takes care of antisymmetrization. 
To compute the isospin part of this matrix element, one writes the baryon 
content of the ket as $\roundket{\Lambda\, N} = |\frac{1}{2}\, m_{t_\Lambda}\; 
\frac{1}{2}\, m_{t_N})$, where $m_{t_N}$ takes the values $m_{t_p}=1/2$ for 
protons and $m_{t_n}=-1/2$ for neutrons, while, in accordance with the 
isospurion stratagem, one treats the $\Lambda$ as if it corresponded to 
$m_{t_\Lambda}=-1/2$. 
On the bra side, one sets $M_T=m_{t_\Lambda}+m_{t_N}$.
To simplify the spatial integration, one resorts to a Moshinsky transformation 
\cite{Mo59} of the initial $\Lambda N$ system.  
To this end, the shell-model radial wave functions are approximated by those of 
a harmonic oscillator with a length parameter of $b=1.75$~fm, which is an 
average between the values appropriate for a $\Lambda$ and for a nucleon 
\cite{Pa95b}. 
Some useful expressions for the computation of these matrix elements are given 
in Appendix~\ref{nuclear}. 

As done in Ref.~\cite{Ba02} and already stated above, the FS effects are taken 
into account as indicated in Subsection \ref{fse}. 
Another important effect to include due to the relatively 
large momentum transfers involved in nonmesonic decays is that of short range 
correlations (SRC). The most satisfactory way to deal with the SRC between the 
$\Lambda$ and the inducing nucleon in the initial state would be through a 
finite-nucleus G-matrix calculation \cite{Ha93}. However, as mentioned in 
Ref.~\cite{Pa97}, this can be well simulated by means of the correlation 
function 
\begin{equation}
g_{\Lambda N}(r) = \left( 1 - e^{-r^2/\alpha^2} \right)^2 + 
\beta r^2\,e^{-r^2/\gamma^2} \,, 
\end{equation}
with $\alpha=0.5$ fm, $\beta=0.25$ fm$^{-2}$ and $\gamma=1.28$ fm. As for the 
SRC between the two emitted nucleons, one might want to perform a T-matrix 
calculation including final state interactions along the lines of 
Ref.~\cite{Ha70}.%
\footnote{In fact, there are claims that this is very important for a good 
description of the nonmesonic decay observables \cite{Pa01}.} 
A simpler, if less satisfactory, way is to again appeal to a correlation 
function, like \cite{We77} 
\begin{equation}
g_{NN}(r) = 1 - j_0(q_c r) \,,
\end{equation}
where $j_0$ is a spherical Bessel function and $q_c=3.93$ fm$^{-1}$. For our 
purposes here, it is sufficient to follow Ref.~\cite{Ba02} and opt for these 
phenomenological correlation functions. Thus, in the calculation of the nuclear 
matrix elements in Eq.~\Ref{matrix} we simply make the replacements 
\begin{eqnarray}
\roundket{j_1 \Lambda \; j_2 N\, J} &\to& 
g_{\Lambda N}(r)\roundket{j_1 \Lambda \; j_2 N\, J} \,, 
\nonumber \\
\roundbra{p'l'P'L'\lambda'S'J'T'M'_T} &\to& 
\roundbra{p'l'P'L'\lambda'S'J'T'M'_T}\, g_{NN}(r) \,. 
\label{src}
\end{eqnarray}

%
%
\settoheight{\thisheight}{ $ \displaystyle B^T_\rho $ }
\begin{table}[!ht] 
\vspace{.5\baselineskip}
\begin{center}
\caption{\label{local} 
Corrections due to the kinematical effects on the nonmesonic decay rates of 
$^{12}_{\,\;\Lambda}$C in several OME models, when only the \emph{local} 
potential is included in the calculation. 
See text for detailed explanation.}
\vspace{.5\baselineskip}

\begin{tabular*}{\textwidth}{@{\extracolsep{\fill}}|crrrrrr|} 
\hline
\hline 

Model/Kinematics
&\cpos{$\Gamma_n^{PC}$}&\cpos{$\Gamma_n^{PV}$}&\cpos{$\Gamma_p^{PC}$}
&\cpos{$\Gamma_p^{PV}$}&\cpos{$\Gamma_{nm}$}&\cposb{
$\Gamma_n/\Gamma_p$}\\
\hline
$\pi$& \rule{0cm}{1.5\thisheight} 
& & & & & \\
averaged\rule{0cm}{1.5\thisheight}
&\none&\none&\none&\none&\none&\noneb\\
$\mu_\pi \rightarrow \tilde{\mu}_\pi$  
&0.0019&0.0230&0.0974&0.0601&0.1823&0.0031\\
full   
&0.0019&0.0224&0.1000&0.0586&0.1829&0.0024\\
\hline 
$(\pi,\eta,K)$ & \rule{0cm}{1.5\thisheight}
& & & & & \\
averaged\rule{0cm}{1.5\thisheight}
&\none&\none&\none&\none&\none&\noneb\\
$\mu_\pi \rightarrow \tilde{\mu}_\pi$   
&0.0024&0.0292&0.0615&0.0653&0.1583&$-0.0275$\\
$\mu_i \rightarrow \tilde{\mu}_i$ 
&0.0023&0.0359&0.0498&0.0707&0.1586&$-0.0150$\\
full  
&0.0023&0.0353&0.0508&0.0694&0.1577&$-0.0156$\\
\hline
$\pi+\rho$& \rule{0cm}{1.5\thisheight}
& & & & & \\
averaged\rule{0cm}{1.5\thisheight}
&\none&\none&\none&\none&\none&\noneb\\
$\mu_\pi \rightarrow \tilde{\mu}_\pi$  
&0.0008&0.0214&0.0712&0.0658&0.1593&0.0055\\
$\mu_i \rightarrow \tilde{\mu}_i$
&0.0008&0.0209&0.0711&0.0685&0.1614&0.0047\\
full   
&0.0008&0.0206&0.0729&0.0652&0.1597&0.0046\\
\hline
$(\pi,\eta,K)+(\rho,\omega,K^*)$ & \rule{0cm}{1.5\thisheight}
& & & & & \\
averaged\rule{0cm}{1.5\thisheight}
&\none&\none&\none&\none&\none&\noneb\\
$\mu_\pi \rightarrow \tilde{\mu}_\pi$   
&0.0048&0.0245&0.0710&0.0803&0.1807&$-0.0103$\\
$\mu_i \rightarrow \tilde{\mu}_i$ 
&0.0058&0.0323&0.0605&0.0953&0.1940&$-0.0032$\\
full 
&0.0058&0.0306&0.0619&0.0878&0.1862&$-0.0033$\\
\hline\hline 
\end{tabular*} 
\end{center}
\end{table}
%
%

\vspace{.5\baselineskip}

Following the discussion in Subsections \ref{nstr} and \ref{str}, we initially 
focus our attention on the reduction of the effective meson masses, 
$\tilde{\mu}_i$ in Eq.~\Ref{emm}, especially that of the pion, $\tilde{\mu}_\pi$ 
in Eq.~\Ref{cf3}, and show that indeed this is the main kinematical effect for 
the local potential, but not so for the nonlocal one. 
To this end we give, in Tables \ref{local} and \ref{nonlocal}, the corrections 
that should be added, according to several different calculations, to the 
standard OME results, \ie, those obtained when both the kinematical and the 
nonlocality effects are completely ignored. 
In Table~\ref{local}, are the corrections corresponding to calculations that  
use only the local potential, and in Table~\ref{nonlocal}, those corresponding 
to calculations that include also the first-order nonlocality terms. In each 
table, the first column indicates which mesons have been included in the 
exchange process and how far the kinematical effects due to the 
lambda-nucleon mass difference have been taken into consideration. The entry 
``averaged'' means that the mass-averaging approximation \Ref{len} has been made 
and, consequently, no kinematical effects have been included, while 
$\mu_\pi \rightarrow \tilde{\mu}_\pi$ or $\mu_i \rightarrow \tilde{\mu}_i$ 
indicates that 
they partly have been, through these replacements made, respectively, for the 
pion alone or for all the mesons, in the expressions for the mass-averaged 
potentials  $\bar{V}^{(0)}_i$ and $\hat{\bar{V}}^{(1)}_i$ in Subsections 
\ref{nstr} and \ref{str}. (Excluding, of course, the factor $G_F \, \mu_\pi^2$.) 
Finally, the entry ``full'' means that the kinematical effects have been fully 
taken into account, by making use of the complete expressions for $V^{(0)}_i$ 
and $\hat{V}^{(1)}_i$ when constructing the local transition potential and (for 
Table~\ref{nonlocal}) its first-order nonlocality correction.

%
%
\settoheight{\thisheight}{ $ \displaystyle B^T_\rho $ }
\begin{table}[ht!]  
\vspace{.5\baselineskip}
\begin{center}
\caption{\label{nonlocal} 
First-order nonlocality corrections for the nonmesonic decay rates of 
$^{12}_{\,\;\Lambda}$C in several OME models, and for different treatments of 
the kinematical effects. 
See text for detailed explanation.}
\vspace{.5\baselineskip}

\begin{tabular*}{\textwidth}{@{\extracolsep{\fill}}|crrrrrr|} 
\hline
\hline 
Model/Kinematics
&\cpos{$\Gamma_n^{PC}$}&\cpos{$\Gamma_n^{PV}$}&\cpos{$\Gamma_p^{PC}$}
&\cpos{$\Gamma_p^{PV}$}&\cpos{$\Gamma_{nm}$}&\cposb{
$\Gamma_n/\Gamma_p$}\\
\hline
$\pi$& \rule{0cm}{1.5\thisheight} 
& & & & & \\
averaged\rule{0cm}{1.5\thisheight}
&\none&\none&\none&\none&\none&\noneb\\
$\mu_\pi \rightarrow \tilde{\mu}_\pi$  
&\none&\none&\none&\none&\none&\noneb\\
full   
&$0.0032$&\none&$0.0729$&\none&$0.0761$&$-0.0062$\\
\hline
 
$(\pi,\eta,K)$ & \rule{0cm}{1.5\thisheight}
& & & & & \\
averaged\rule{0cm}{1.5\thisheight} 
&\none&\none&\none&\none&\none&\noneb\\
$\mu_\pi \rightarrow \tilde{\mu}_\pi$  
&\none&\none&\none&\none&\none&\noneb\\
$\mu_i \rightarrow \tilde{\mu}_i$ 
&\none&\none&\none&\none&\none&\noneb\\
full  
&$0.0037$&$-0.0332$&$0.0312$&$-0.0335$&$-0.0318$&$-0.0401$\\
\hline
$\pi+\rho$& \rule{0cm}{1.5\thisheight}
& & & & & \\
averaged\rule{0cm}{1.5\thisheight} 
&$0.0001$&$0.0016$&$0.0004$&$-0.0102$&$-0.0080$&$0.0034$\\
$\mu_\pi \rightarrow \tilde{\mu}_\pi$   
&$0.0001$&$0.0017$&$0.0003$&$-0.0110$&$-0.0088$&$0.0032$\\
$\mu_i \rightarrow \tilde{\mu}_i$
&$0.0001$&$0.0018$&$0.0003$&$-0.0114$&$-0.0091$&$0.0033$\\
full   
&$0.0023$&$0.0014$&$0.0386$&$-0.0088$&$0.0334$&$-0.0001$\\
\hline
$(\pi,\eta,K)+(\rho,\omega,K^*)$ & \rule{0cm}{1.5\thisheight}
& & & & & \\
averaged\rule{0cm}{1.5\thisheight} 
&0.0013&$-0.0471$&$-0.0008$&$-0.0970$&$-0.1435$&$-0.0238$\\
$\mu_\pi \rightarrow \tilde{\mu}_\pi$  
&0.0014&$-0.0500$&$-0.0011$&$-0.1027$&$-0.1524$&$-0.0227$\\
$\mu_i \rightarrow \tilde{\mu}_i$ 
&0.0015&$-0.0522$&$-0.0011$&$-0.1071$&$-0.1589$&$-0.0229$\\
full 
&0.0133&$-0.0901$&0.0404&$-0.1425$&$-0.1790$&$-0.0515$\\
\hline\hline 
\end{tabular*} 
\end{center}
\end{table}
%
%
  
Examining the first block in Table~\ref{local}, one notices immediately that, 
when only the local potential is included in the calculation, the kinematical 
effects are well represented, for the pion, by the replacement $\mu_\pi \to 
\tilde{\mu}_\pi$ in the expression \Ref{avV0pic} for the local transition 
potential obtained when they are completely negleted. 
Thus, the further modifications caused by these effects  in the local potential, 
which lead to the ``full'' expression \Ref{fullV0pic}, are of less importance. 
Comparing the last two lines in the remaining blocks of this table, one 
concludes that the analogous statement holds also for the other mesons. Finally, 
knowing this and comparing the second and third lines in these same blocks, one 
can see that the main local kinematical correction is that affecting the pion 
exchange. 
All these conclusions are in agreement with the discussion in Subsections 
\ref{pion}, \ref{rho}, \ref{etaomega}, \ref{kaon} and \ref{kstar}.

%
%
\settoheight{\thisheight}{ $ \displaystyle B^T_\rho $ }
\begin{table}[!ht]  
\vspace{.5\baselineskip}
\begin{center}
\caption{\label{analysis}
Analysis of the different contributions to the nonmesonic decay rates of 
$^{12}_{\,\;\Lambda}$C in several OME models. All corrections are computed with 
``full'' kinematics. 
See text for detailed explanation.}
\vspace{.5\baselineskip}

\begin{tabular*}{\textwidth}{@{\extracolsep{\fill}}|crrrrrr|} 
\hline
\hline 

Model/Contributions 
&\cpos{$\Gamma_n^{PC}$}&\cpos{$\Gamma_n^{PV}$}&\cpos{$\Gamma_p^{PC}$}
&\cpos{$\Gamma_p^{PV}$}&\cpos{$\Gamma_{nm}$}&\cposb{
$\Gamma_n/\Gamma_p$}\\
\hline
$\pi$ 
& \rule{0cm}{1.5\thisheight} & & & & & \\ 
\emph{Uncorrected value}\rule{0cm}{1.5\thisheight}
&0.0063&0.1162&0.6019&0.2887&1.0131&0.1375\\
\emph{Local kinem. corr.}  
&0.0019&0.0224&0.1000&0.0586&0.1829&0.0024\\
\emph{1$^{\mathit{st}}$-order nonloc. corr.} 
&$0.0032$&\none&$0.0729$&\none&$0.0761$&$-0.0062$\\
\emph{Corrected value}  
&0.0114&0.1386&0.7748&0.3473&1.2721&0.1337\\
\hline
 
$(\pi,\eta,K)$ 
& \rule{0cm}{1.5\thisheight} & & & & & \\
\emph{Uncorrected value}\rule{0cm}{1.5\thisheight}
&0.0056&0.2345&0.2124&0.3813&0.8339&0.4045\\
\emph{Local kinem. corr.}   
&0.0023&0.0353&0.0508&0.0694&0.1577&$-0.0156$\\
\emph{1$^{\mathit{st}}$-order nonloc. corr.} 
&$0.0037$&$-0.0332$&$0.0312$&$-0.0335$&$-0.0318$&$-0.0401$\\
\emph{Corrected value}  
&0.0116&0.2366&0.2944&0.4172&0.9598&0.3488\\
\hline
$\pi+\rho$
& \rule{0cm}{1.5\thisheight} & & & & & \\
\emph{Uncorrected value}\rule{0cm}{1.5\thisheight}
&0.0056&0.1004&0.5140&0.3608&0.9807&0.1212\\
\emph{Local kinem. corr.}    
&0.0008&0.0206&0.0729&0.0652&0.1597&0.0046\\
\emph{1$^{\mathit{st}}$-order nonloc. corr.}
&$0.0023$&$0.0014$&$0.0386$&$-0.0088$&$0.0334$&$-0.0001$\\
\emph{Corrected value}   
&0.0087&0.1224&0.6255&0.4172&1.1738&0.1257\\
\hline
$(\pi,\eta,K)+(\rho,\omega,K^*)$
& \rule{0cm}{1.5\thisheight} & & & & & \\
\emph{Uncorrected value}\rule{0cm}{1.5\thisheight}
&0.0241&0.2218&0.2961&0.6238&1.1657&0.2672\\
\emph{Local kinem. corr.}   
&0.0058&0.0306&0.0619&0.0878&0.1862&$-0.0033$\\
\emph{1$^{\mathit{st}}$-order nonloc. corr.} 
&0.0133&$-0.0901$&0.0404&$-0.1425$&$-0.1790$&$-0.0515$\\
\emph{Corrected value} 
&0.0432&0.1623&0.3984&0.5691&1.1729&0.2124\\
\hline\hline 
\end{tabular*} 
\end{center}
\end{table}
%
%

Going now to Table~\ref{nonlocal}, one sees that the situation is quite 
different as regards the influence of the kinematical effects on the nonlocal  
potential. In fact, the first two blocks show that, in OME models 
containing only pseudoscalar mesons, the first-order nonlocality corrections 
vanish unless one takes the kinematical effects fully into account. This is just 
a restatement of Eqs. \Ref{avV1pic}, \Ref{avV1kaonc} and the analogous result 
for the $\eta$ meson. 
Similarly, examination of the last two blocks shows that, in OME models 
containing vector mesons, if one does not take the kinematical effects 
fully into account  the first-order nonlocality corrections generally turn out  
very different from their actual values. 
The mere replacement $\mu_i \to \tilde{\mu}_i$ does not work well in this case.  
This is so because, as can be seen in Subsections \ref{rho} and \ref{kstar}, 
several nonlocal terms appear as a direct consequence of the kinematical 
effects, rather than simply being modified by them. Therefore, to be 
consistent, one should take the kinematical effects due to the lambda-nucleon 
mass difference \emph{fully} into account when dealing with the nonlocality 
corrections. 

To better visualize our findings, we exhibit in Table~\ref{analysis} an analysis 
of the different contributions to the nonmesonic decay rates of 
$^{12}_{\,\;\Lambda}$C in the four OME models we have been considering. 
In the first line of each block, we give the values that would be obtained for 
the transition rates in the standard OME approach, \ie, when neither the 
kinematical, nor the nonlocality corrections are included. 
On the second line, are the corrections to these values arising from the 
kinematical effects related to the lambda-nucleon mass difference, but still 
restricted to the local contributions only. 
On the third line, we have the first-order nonlocality corrections, and on the 
last one, the values of the decay rates including the two corrections.  
As required by consistency, according to our previous discussion, both 
corrections are computed with the kinematical effects fully taken into account. 
  
Examining this table, one notices that the kinematical and the nonlocality 
corrections are typically of comparable  sizes. 
Furthermore, for the partial and total decay rates, the former ones are always 
positive, while the latter are sometimes negative. 
Consequently, the two corrections should be included simultaneously, or not at 
all.
Another point to remark is that the modifications in the uncorrected values of 
these decay rates when going from one OME model to another are of the same 
general magnitude as these corrections within each model. Therefore, 
it might be questionable to consider other mesons besides the pion 
without, at the same time, including the kinematical and nonlocality 
corrections. 

The influence of the two effects together in the several partial decay rates 
varies around $\sim 80\%$ for $\Gamma^{PC}_n$ and $\sim 20\%$ for the other 
ones, depending on the OME model. The net effect on the main decay obervables is 
smaller: it is $\sim 15\%$ for $\Gamma_{nm}$ and $\sim 10\%$ for the ratio  
$\Gamma_n/\Gamma_p$, again depending on the OME model considered. As one can 
see, these corrections are of no help to solve the discrepancy between the 
theoretical prediction and the experimental determinations for the latter 
quantity. They are too small for that, and usually go in the wrong direction.
As a final observation, notice that the combined correction affects very 
differently the parity-conserving and the parity-violating transitions, 
especially when strange mesons are involved. 
For instance, in the model including the full pseudoscalar 
and vector meson octets (fourth block), the uncorrected value for 
$\Gamma^{PC}/\Gamma^{PV}$ is 0.379, while the corrected one is 0.604.

%
%
%
\settoheight{\thisheight}{ $ \displaystyle B^T_\rho $ }
\begin{table}[!ht]  
\vspace{.5\baselineskip}
\begin{center}
\caption{\label{singlemesons}
Analysis of the different contributions to the nonmesonic decay rates of 
$^{12}_{\,\;\Lambda}$C coming from each meson acting alone. The interference 
among the different mesons is ignored. All corrections are computed with 
``full'' kinematics. See text for detailed explanation.}
\vspace{.5\baselineskip}

\begin{tabular*}{\textwidth}{@{\extracolsep{\fill}}|crrrrrr|} 
\hline
\hline 

Meson/Contributions 
&\cpos{$\Gamma_n^{PC}$}&\cpos{$\Gamma_n^{PV}$}&\cpos{$\Gamma_p^{PC}$}
&\cpos{$\Gamma_p^{PV}$}&\cpos{$\Gamma_{nm}$}&\cposb{
$\Gamma_n/\Gamma_p$}\\
\hline
$\pi$ 
& \rule{0cm}{1.5\thisheight} & & & & & \\ 
\emph{Uncorrected value}\rule{0cm}{1.5\thisheight}
&0.0063&0.1162&0.6019&0.2887&1.0131&0.1375\\
\emph{Local kinem. corr.}  
&0.0019&0.0224&0.1000&0.0586&0.1829&0.0024\\
\emph{1$^{\mathit{st}}$-order nonloc. corr.} 
&$0.0032$&\none&$0.0729$&\none&$0.0761$&$-0.0062$\\
\emph{Corrected value}  
&0.0114&0.1386&0.7748&0.3473&1.2721&0.1337\\
\hline
 
$\eta$ 
& \rule{0cm}{1.5\thisheight} & & & & & \\
\emph{Uncorrected value}\rule{0cm}{1.5\thisheight}
&0.0020&0.0031&0.0047&0.0024&0.0122&0.7200\\
\emph{Local kinem. corr.}   
&\none &0.0002&0.0002&0.0001&0.0006&$-0.0047$\\
\emph{1$^{\mathit{st}}$-order nonloc. corr.} 
&$0.0003$& \none  &$0.0005$& \none  &$0.0007$&$-0.0130$\\
\emph{Corrected value}  
&0.0023&0.0033&0.0054&0.0025&0.0135&0.7023\\
\hline
$K$
& \rule{0cm}{1.5\thisheight} & & & & & \\
\emph{Uncorrected value}\rule{0cm}{1.5\thisheight}
&0.0058&0.0453&0.0780&0.0277&0.1569&0.4841\\
\emph{Local kinem. corr.}    
&0.0002&0.0041&0.0061&0.0025&0.0129&0.0009\\
\emph{1$^{\mathit{st}}$-order nonloc. corr.}
&$-0.0002$&$-0.0233$&$0.0073$&$-0.0139$&$-0.0303$&$-0.1886$\\
\emph{Corrected value}   
&0.0058&0.0261&0.0914&0.0163&0.1395&0.2964\\
\hline
$\rho$
& \rule{0cm}{1.5\thisheight} & & & & & \\
\emph{Uncorrected value}\rule{0cm}{1.5\thisheight}
&0.0016&0.0017&0.1274&0.0082&0.1390&0.0246\\
\emph{Local kinem. corr.}    
&0.0001&0.0002&0.0070&0.0003&0.0075&0.0004\\
\emph{1$^{\mathit{st}}$-order nonloc. corr.}
&$-0.0005$&$0.0007$&$-0.0069$&$-0.0013$&$-0.0080$&$0.0029$\\
\emph{Corrected value}   
&0.0012&0.0026&0.1275&0.0072&0.1385&0.0279\\
\hline
$\omega$
& \rule{0cm}{1.5\thisheight} & & & & & \\
\emph{Uncorrected value}\rule{0cm}{1.5\thisheight}
&0.0073&0.0020&0.0732&0.0019&0.0843&0.1232\\
\emph{Local kinem. corr.}    
&0.0007&0.0001&0.0051&0.0001&0.0062&0.0028\\
\emph{1$^{\mathit{st}}$-order nonloc. corr.}
&$0.0016$&$0.0037$&$-0.0030$&$0.0011$&$0.0034$&$0.0708$\\
\emph{Corrected value}   
&0.0096&0.0058&0.0753&0.0031&0.0939&0.1968\\
\hline
$K^*$
& \rule{0cm}{1.5\thisheight} & & & & & \\
\emph{Uncorrected value}\rule{0cm}{1.5\thisheight}
&0.0062&0.0145&0.0614&0.0189&0.1010&0.2579\\
\emph{Local kinem. corr.}   
&0.0001&0.0011&0.0021&0.0007&0.0040&$0.0060$\\
\emph{1$^{\mathit{st}}$-order nonloc. corr.} 
&$-0.0016$&$0.0086$&$-0.0025$&$-0.0004$&$0.0041$&$0.0960$\\
\emph{Corrected value} 
&0.0047&0.0242&0.0610&0.0192&0.1091&0.3599\\
\hline\hline 
\end{tabular*} 
\end{center}
\end{table}
%
%

One might wonder how important the corrections are for each meson-exchange 
taken in isolation. To answer this question, we show in Table~\ref{singlemesons} 
the different contributions to the partial and total decay rates, as well as to 
the $n/p$ ratio, coming from each meson acting alone. Of course, in actual fact  
the contributions of the several mesons interfere with each other, so that the 
numbers shown in this table do not have any direct physical meaning, but they 
serve as an indication of the importance of the two corrections for each meson. 

The local kinematical corrections affect the pion more than any other meson, as 
expected due to its low mass. There, the effect on the decay rates is of the 
order of 20 -- 30\%, while for the other mesons it does not exceed $\sim 10\%$. 
For the $n/p$ ratio, the effect is always negligible. 

The first-order nonlocality corrections vary a lot, both in sign, and in 
magnitude. Depending on the meson and on the partial rate considered, the 
corrections can be as low as a few per cent, but in many cases reach the 
$50\%$ level. 
The effect on the omega-exchange is exceptionally large, specially for the 
parity-violating transitions, where, in fact, more than half of the contribution 
comes from the nonlocal terms. However, for this meson the parity-conserving 
transitions are far more important and, besides this, the net nonlocal $PC$  
contribution is opposite to the $PV$ one, such that the final effect of the 
nonlocal corrections on the total decay rate is of less than 5\%. This is again 
an indication that the effects we are considering here are not exactly small, 
although, due to a series of compensations, they end up by not affecting the 
usual decay observables too much. 

\subsection{Asymmetry parameter} \label{asymmetry}  

The only remaining nonmesonic decay observable, beyond $\Gamma_{nm}$ and 
$\Gamma_n/\Gamma_p$, that has been measured to date is the asymmetry parameter, 
$a_\Lambda$, which depends on the interference between the amplitudes for $PC$ 
and $PV$ proton-induced transitions to final states with different isospins. 
This parameter is a characteristic of the nonmesonic decay of a polarized 
$\Lambda$ in the nuclear medium, having been defined so as to subdue the 
influence of the particular hypernucleus considered \cite{Ra92}. 
It is experimentally extracted from measurements of the asymmetry, $A_y$, in the 
angular distribution of the emitted protons in the nonmesonic decay of polarized 
hypernuclei \cite{Aj92,Aj00}. 
There are large discrepancies, both experimentally and theoretically, in the 
determination of  $a_\Lambda$ \cite[\emph{Sec.7}]{Al02}, specially after the 
newest experimental results for the decay of $^5_\Lambda\mathrm{He}$  
\cite{Aj00}, which give a positive value for this observable, differently from 
all previous measurements. 
In strong disagreement with that, all calculations so far find a negative value 
for $a_\Lambda$ \cite{Al02}, which makes the investigation of the kinematical 
plus nonlocality corrections on this observable particularly relevant. 

For the case of $^5_\Lambda\mathrm{He}$, the following expression can be used 
to compute the asymmetry parameter \cite{Al02}:
\begin{equation}\label{asym}
a_\Lambda = A_y(^5_\Lambda \mathrm{He}) = 
\frac{2\,\Re \left[\sqrt{3}\, ae^* - b\, ( c^* - \sqrt{2}\, d^* ) + 
\sqrt{3}\, f\, ( \sqrt{2}\, c^* + d^* ) \right]}
{ |a|^2 + |b|^2 + 3 \left( |c|^2 + |d|^2 + |e|^2 + |f|^2 \right) } \,,
\end{equation}
where
\settoheight{\thisheight}{$\displaystyle 2$} 
\begin{eqnarray}
a &=& \bra{np, ^1\!\mathrm{S}_0}\hat{V}\ket{\Lambda p, ^1\!\mathrm{S}_0}
\nonumber \\ &=&  
\mathcal{M}(pl\!=\!0 \; PL\!=\!0 \; \lambda\!=\!0 \; S\!=\!0 \; J\!=\!0 \; 
T\!=\!1 \; M_T\!=\!0 \; ; \Lambda p, (1s_{1/2})^2\, J\!=\!0 ) 
\,, 
\rule[-1\thisheight]{0cm}{1\thisheight}   \qquad
\nonumber \\  
b &=& i \bra{np, ^3\!\mathrm{P}_0}\hat{V}\ket{\Lambda p, ^1\!\mathrm{S}_0}
\nonumber \\ &=& 
i \mathcal{M}(pl\!=\!1 \; PL\!=\!0 \; \lambda\!=\!1 \; S\!=\!1 \; J\!=\!0 \; 
T\!=\!1 \; M_T\!=\!0 \; ; \Lambda p, (1s_{1/2})^2\, J\!=\!0 ) 
\,,   \rule[-1\thisheight]{0cm}{1\thisheight}   \qquad
\nonumber \\ 
c &=& \bra{np, ^3\!\mathrm{S}_1}\hat{V}\ket{\Lambda p, ^3\!\mathrm{S}_1}
\nonumber \\ &=& 
- \mathcal{M}(pl\!=\!0 \; PL\!=\!0 \; \lambda\!=\!0 \; S\!=\!1 \; J\!=\!1 \; 
T\!=\!0 \; M_T\!=\!0 \; ; \Lambda p, (1s_{1/2})^2\, J\!=\!1 ) 
\,,   \rule[-1\thisheight]{0cm}{1\thisheight}   \qquad 
\nonumber \\ 
d &=& - \bra{np, ^3\!\mathrm{D}_1}\hat{V}\ket{\Lambda p, ^3\!\mathrm{S}_1}
\nonumber \\ &=& 
 \mathcal{M}(pl\!=\!2 \; PL\!=\!0 \; \lambda\!=\!2 \; S\!=\!1 \; J\!=\!1 \; 
T\!=\!0 \; M_T\!=\!0 \; ; \Lambda p, (1s_{1/2})^2\, J\!=\!1 ) 
\,,   \rule[-1\thisheight]{0cm}{1\thisheight}   \qquad 
\nonumber \\ 
e &=& i \bra{np, ^1\!\mathrm{P}_1}\hat{V}\ket{\Lambda p, ^3\!\mathrm{S}_1}
\nonumber \\ &=& 
- i \mathcal{M}(pl\!=\!1 \; PL\!=\!0 \; \lambda\!=\!1 \; S\!=\!0 \; J\!=\!1 \; 
T\!=\!0 \; M_T\!=\!0 \; ; \Lambda p, (1s_{1/2})^2\, J\!=\!1 ) 
\,,   \rule[-1\thisheight]{0cm}{1\thisheight}   \qquad 
\nonumber \\ 
f &=& - i \bra{np, ^3\!\mathrm{P}_1}\hat{V}\ket{\Lambda p, ^3\!\mathrm{S}_1}
\nonumber \\ &=& 
- i \mathcal{M}(pl\!=\!1 \; PL\!=\!0 \; \lambda\!=\!1 \; S\!=\!1 \; J\!=\!1 \; 
T\!=\!1 \; M_T\!=\!0 \; ; \Lambda p, (1s_{1/2})^2\, J\!=\!1 ) 
\,. \qquad 
\label{tramps}
\end{eqnarray}
The extra factors in the transition amplitudes are due to differences in phase 
conventions, as explained in Appendix~\ref{conventions}, and we have rewritten 
them in terms of the nuclear matrix elements defined in Eq.~\Ref{matrix}. 

It is important to realize that the formula \Ref{asym} is not of general 
validity, and is, in fact, merely an approximation taken over from the result 
valid for a free space process \cite{Na99} and adapted somehow to the 
hypernuclear decay situation. 
In particular, there is no definitive prescription on how to devide the energy 
liberated in the decay, $\Delta_F$, between the relative and CM motions of the 
emitted nucleons. 
It seems that, based on the expectation that the final result be insensitive to 
this point, some authors merely take $P=0$ in Eqs.~\Ref{tramps}, while others 
integrate over phase space, both the numerator, and the denominator in 
Eq.~\Ref{asym}. 
We have checked that the two prescriptions indeed give similar results for 
$^5_\Lambda \mathrm{He}$, and opted to tabulate only those corresponding to the 
first one. 
A more rigorous calculation of $a_\Lambda$ is planned for the near future.

%
%
\settoheight{\thisheight}{ $ \displaystyle B^T_\rho $ }
\begin{table}[!ht]  
\vspace{.5\baselineskip}
\begin{center}
\caption{\label{helium}
Analysis of the different contributions to the nonmesonic decay rates and 
asymmetry parameter of $^5_\Lambda \mathrm{He}$ in several OME models. 
All corrections are computed with ``full'' kinematics. 
See text for detailed explanation.}
\vspace{.5\baselineskip}

\begin{tabular*}{\textwidth}{@{\extracolsep{\fill}}|crrrrrr|} 
\hline
\hline 

Model/Contributions 
&\cpos{$\Gamma_n^{PC}$}&\cpos{$\Gamma_n^{PV}$}&\cpos{$\Gamma_p^{PC}$}
&\cpos{$\Gamma_p^{PV}$}&\cpos{$\Gamma^{PC}/\Gamma^{PV}$}&\cposb{
$a_\Lambda$}\\
\hline
$\pi$ 
& \rule{0cm}{1.5\thisheight} & & & & & \\ 
\emph{Uncorrected value}\rule{0cm}{1.5\thisheight}
&0.0004&0.0739&0.3889&0.1479&$1.7553$&$-0.4351$\\
\emph{Local kinem. corr.}  
&0.0005&0.0130&0.0599&0.0258&$-0.0295$&$-0.0084$\\
\emph{1$^{\mathit{st}}$-order nonloc. corr.} 
&$0.0011$&\none&$0.0461$&\none&$0.1810$&$-0.0021$\\
\emph{Corrected value}  
&$0.0020$&$0.0869$&$0.4949$&$0.1737$&$1.9068$&$-0.4456$\\
\hline
 
$(\pi,\eta,K)$ 
& \rule{0cm}{1.5\thisheight} & & & & & \\
\emph{Uncorrected value}\rule{0cm}{1.5\thisheight}
&$0.0013$&$0.1734$&$0.1261$&$0.2077$&$0.3342$&$-0.5852$\\
\emph{Local kinem. corr.}   
&$0.0008$&$0.0234$&$0.0280$&$0.0327$&$0.0229$&$-0.0106$\\
\emph{1$^{\mathit{st}}$-order nonloc. corr.} 
&$0.0013$&$-0.0285$&$0.0176$&$-0.0215$&$0.0952$&$-0.0406$\\
\emph{Corrected value}  
&$0.0034$&$0.1683$&$0.1717$&$0.2189$&$0.4523$&$-0.6364$\\
\hline
$\pi+\rho$
& \rule{0cm}{1.5\thisheight} & & & & & \\
\emph{Uncorrected value}\rule{0cm}{1.5\thisheight}
&$0.0001$&$0.0648$&$0.3135$&$0.1939$&$1.2118$&$-0.2665$\\
\emph{Local kinem. corr.}    
&$0.0001$&$0.0121$&$0.0433$&$0.0299$&$-0.0251$&$-0.0217$\\
\emph{1$^{\mathit{st}}$-order nonloc. corr.}
&$0.0004$&$0.0016$&$0.0245$&$-0.0056$&$0.1003$&$-0.0273$\\
\emph{Corrected value}   
&$0.0006$&$0.0785$&$0.3813$&$0.2182$&$1.2870$&$-0.3155$\\
\hline
$(\pi,\eta,K)+(\rho,\omega,K^*)$
& \rule{0cm}{1.5\thisheight} & & & & & \\
\emph{Uncorrected value}\rule{0cm}{1.5\thisheight}
&$0.0112$&$0.1672$&$0.1778$&$0.3642$&$0.3558$&$-0.5131$\\
\emph{Local kinem. corr.}   
&$0.0026$&$0.0206$&$0.0344$&$0.0442$&$0.0233$&$-0.0090$\\
\emph{1$^{\mathit{st}}$-order nonloc. corr.} 
&$0.0054$&$-0.0715$&$0.0237$&$-0.0897$&$0.2073$&$-0.0167$\\
\emph{Corrected value} 
&$0.0192$&$0.1163$&$0.2359$&$0.3187$&$0.5864$&$-0.5388$\\
\hline\hline 
\end{tabular*} 
\end{center}
\end{table}
%
%

We give, in Table~\ref{helium}, the results we obtained for the decay 
rates and asymmetry parameter of $^5_\Lambda \mathrm{He}$. 
The calculation goes along similar lines to those explained in 
Subsection~\ref{rates}, except for the obvious changes. 
The (hyper)nuclear model-space is restricted to the $1s_{1/2}$ orbital; 
the relevant nuclear structure factors, Eq.~\Ref{structure}, are \cite{Kr02} 
$F^{1s_{1/2}\,n}_{0} = F^{1s_{1/2}\,p}_{0} = 1/2$, $F^{1s_{1/2}\,n}_{1} = 
F^{1s_{1/2}\,p}_{1} = 3/2$; we used $1.62$~fm for the oscillator length 
parameter; and took $\Delta_F = 153.83$~MeV. 
It is clear from this table that the comments made above about the decay rates 
of $^{12}_{\,\;\Lambda}$C remain qualitatively valid in the present case. 
As to the asymmetry parameter, the effect of the two corrections reaches 
$\sim 18\%$ in the $\pi + \rho$ model, but is of only $\sim 5\%$ in the complete 
model. On the average, it varies around $\sim 10\%$. 
It is interesting to observe that the two corrections always go in the direction 
of making $a_\Lambda$ even more negative, thus confirming the sign of the  
existing theoretical predictions for this observable in contraposition to its 
most recent experimental determination.

\section{Summary and conclusions} \label{con}

We have proposed an approach that naturally establishes a hierarchy for the 
different levels of approximation in the extraction of the nonrelativistic 
transition potential in OME models in general. The central result is 
Eq.~\Ref{Vmom}. The first term corresponds to the local approximation, usually 
adopted in the literature on nonmesonic decay. The second one, to the 
first-order nonlocality correction, which we have included in our calculations 
here. And the last one, to the second-order nonlocality correction, which we 
have neglected. We have also given a detailed and general account on how to deal 
accurately with the kinematical effects that result when one has different 
baryon masses on the four legs in the OME Feynman amplitude in Fig.~\ref{ampc}. 
All this was particularized to $\Lambda$ hypernuclear nonmesonic decay and 
detailed expressions for all contributions to the transition potential coming 
from the exchange of the complete pseudoscalar and vector meson octets were 
given.

Using this formalism, we have investigated the relative importance of two 
effects sistematically ignored in OME models for the nonmesonic weak decay of 
$\Lambda$-hypernuclei.  
First, that of an accurate treatment of the kinematics, \ie, of taking into 
account the difference in mass between the hyperon and the nucleon, when 
determining the OME transition potential. 
Secondly, we considered the influence of the first-order nonlocality-correction 
terms.  
Surprisingly, in view of the nonnegligible value of the mass-asymmetry ratio in 
Eq.~\Ref{ldn}, we came to the conclusion that the kinematical effect on the 
local potential is small, except for the reduction of the effective mass of the 
pion, Eq.~\Ref{cf3}, which implies an increase
 of $\sim 35\%$ in the range of 
the corresponding transition potential. However its indirect influence is 
important, since it activates several nonlocal terms in the transition 
potential. 

Our conclusion is that the influence of the two effects together on the partial 
decay rates is sizeable, the full amount depending on which mesons are 
included. It can be very large for $\Gamma^{PC}_n$, often exceeding $100\%$, 
while it typically stays in the  $20$--$30\%$ range for the other partial rates. 
The effects are somewhat washed out in the main decay observables, averaging to  
$\sim 15\%$ for the total nonmesonic decay rate, 
$\Gamma_{nm}=\Gamma_n+\Gamma_p$, and to $\sim 10\%$ for, both the neutron-  to 
proton-induced ratio, $\Gamma_n/\Gamma_p$, and the asymmetry parameter, 
$a_\Lambda$. 
In particular, they do not in any way account for the well known discrepancy 
between the standard OME predictions for the latter two observables and their 
measured values \cite{Al02}. 
To summarize, although the kinematical and nonlocality effects can be very 
important for particular transitions, they end up by not affecting the main 
decay observables too much. One can partly understand this from the following 
two facts:
\begin{enumerate}
\item
The most affected transitions are by far the neutron-induced, parity-conserving 
ones. However those contribute very little to $\Gamma_{nm}$ and to  
$\Gamma_n/\Gamma_p$, and not at all to $a_\Lambda$, which is a property of 
proton-induced transitions only.
\item
The largest relative effect of the nonlocality corrections is on the $PV$ 
transitions coming from the one-omega-exchange process. However, for this meson, 
the $PC$ transitions are much more important and they are affected in the 
opposite direction. As a result the two effects are largely neutralized in any 
of the main decay observables, which depend on, either the sum of the 
intensities, or the product of the amplitudes for these two types of 
transitions. 
\end{enumerate}

We have also shown that the parameter $\Gamma^{PC}/\Gamma^{PV}$ is strongly 
affected in OME models that include strange mesons. The relative correction is 
of $\sim 30\%$ in the $(\pi+\eta+K)$ model and of $\sim 60\%$ in the complete 
model. However, none of the three nonmesonic decay observables that have been 
measured up to now is sensitive to this ratio.

Let us finalize by saying that, although the effects we have studied here can be 
considered small in view of the imprecision of the available measurements and 
the degree of uncertainty presently existing in the parameters  of OME models 
for nonmesonic decay (particularly coupling constants), they are not altogether 
negligible. 
In fact, in many cases they appear to influence the theoretical predictions by 
roughly as much as the inclusion of other mesons beyond the pion in the exchange 
process. 
It seems, therefore, that they have a part to play in OME models for nonmesonic 
hypernuclear decays, specially if one takes into consideration that more 
detailed and accurate experimental data on this issue is forthcoming. 

%
%

\vspace{2\baselineskip}
\noindent
\textbf{Acknowledgements}
\vspace{1\baselineskip}

C. B. and F. K.  are fellows of the CONICET (Argentina) and acknowledge the 
support of CONICET under grant PIP 463, of ANPCyT (Argentina) under grant BID 
1201/OC-AR (PICT 03-04296) and of Fundaci\'on Antorchas (Argentina). The work of 
C. D. C. was supported by FAPESP (S\~ao Paulo, Brazil). 
One of us (C. B.) would like to thank G. Garbarino, A. Parre\~no and A. Ramos 
for helpful discussions about the asymmetry parameter.

\vspace{2\baselineskip}
\begin{appendix}
\vspace{6ex}

{\noindent \Large \bfseries Appendices}

\section{Calculation of nuclear matrix elements \label{nuclear}}
\appendixeqn

We collect in this appendix some useful expressions for the calculation of the 
several matrix elements contributing to Eq.~\Ref{matrix}. As mentioned below 
that equation, the first step is to factor out the baryon content of the initial 
state,
\begin{equation}
\roundket{j_1 \Lambda \; j_2 N\,J} = \roundket{j_1\, j_2\, J}\; 
\roundket{\Lambda N} \equiv \roundket{n_1 l_1 j_1\, n_2 l_2 j_2\, J}\; 
\roundket{\Lambda N} \,,
\end{equation}
and perform a Moshinsky transformation \cite{Mo59} for its space-spin part,%
\begin{eqnarray} \!\!\!\!\!\!\!\!\!\!\!\!
\roundket{n_1 l_1 j_1\, n_2 l_2 j_2\, J} &=& \sqrt{(2j_1+1)(2j_2+1)}\;\; 
\sum_{\lambda S} \sqrt{(2\lambda + 1)(2S+1)}\, 
\ninej{l_1}{\frac{1}{2}}{j_1}{l_2}{\frac{1}{2}}{j_2}{\lambda}{S}{J} 
\nonumber \\ & & {} \times  \sum_{nlNL} \roundket{nl\,NL\, \lambda S J} \, 
\roundbra{nl\, NL\, \lambda} \left. n_1 l_1\, n_2 l_2\, \lambda \right) \,. 
\label{moshinsky}
\end{eqnarray}
A look at Subsections \ref{nstr} and \ref{str} shows that the operators involved 
are always of the general form $v(r)\, \Omega(\bold{\sigma}_1, \bold{\sigma}_2, 
\hat{\bold{r}}, \nabla)$. 
All matrix elements are diagonal in the quantum numbers $L$ and $J$, and the 
needed results are listed below. 
Some of them have already been given in Ref.~\cite{Ba02} but are repeated here 
for completeness.%
\footnote{There are some misprints in the formulas for the matrix elements in 
the published version of Ref.\cite{Ba02}. These misprints, however, do not occur 
in its preprint version, \texttt{axXiv:nucl-th/0011092}.}
\begin{eqnarray} 
\lefteqn{ 
\roundbra{p'l'\,P'L\,\lambda'S'J}v(r)\roundket{nl\,NL\,\lambda SJ} = 
}
\nonumber \\ & & 
\delta_{l'l} \delta_{\lambda'\lambda} \delta_{S'S} \roundbraket{P'L}{NL} 
\roundbra{p'l'}v(r)\roundket{nl} \,, 
\label{nmei} 
\end{eqnarray}
\begin{eqnarray} 
\lefteqn{ 
\roundbra{p'l'\,P'L\,\lambda'S'J} v(r) \bold{\sigma}_1\cdot\bold{\sigma}_2  
\roundket{nl\,NL\,\lambda SJ} = 
} 
\nonumber \\ & & 
\delta_{l'l} \delta_{\lambda'\lambda} \delta_{S'S} 
\left[ 2S(S+1) - 3 \right]\, 
\roundbraket{P'L}{NL} \roundbra{p'l'}v(r)\roundket{nl} \,, 
\end{eqnarray} 
\begin{eqnarray} 
\lefteqn{ 
\roundbra{p'l'\,P'L\,\lambda'S'J} v(r)S_{12}(\hat{\bold{r}}) 
\roundket{nl\,NL\,\lambda SJ} = 
} 
\nonumber \\ & & 
(-)^{L+J+1}\, \delta_{S1} \delta_{S'1} 
\sqrt{120 (2l'+1)(2l+1)(2\lambda'+1)(2\lambda+1)}\, 
\nonumber \\ & & \times 
\threej{l'}{2}{l}{0}{0}{0} 
\sixj{\lambda'}{2}{\lambda}{l}{L}{l'} 
\sixj{1}{2}{1}{\lambda}{J}{\lambda'} 
\roundbraket{P'L}{NL} \roundbra{p'l'}v(r)\roundket{nl} \,, 
\end{eqnarray} 
\settowidth{\thiswidth}{$ \displaystyle (-)^{S'+S}\, $}
\begin{eqnarray} 
\lefteqn{ 
(-)^{S'+S}\, 
\roundbra{p'l'\,P'L\,\lambda'S'J} v(r)\bold{\sigma}_1 \cdot \hat{\bold{r}}  
\roundket{nl\,NL\,\lambda SJ} = 
} 
\nonumber \\ 
\lefteqn{ 
\rule{1.\thiswidth}{0cm} 
\roundbra{p'l'\,P'L\,\lambda'S'J} v(r)\bold{\sigma}_2 \cdot \hat{\bold{r}}  
\roundket{nl\,NL\,\lambda SJ} = 
}
\nonumber \\ & & 
(-)^{L+J+1}\, 
\sqrt{6 (2l'+1)(2l+1)(2\lambda'+1)(2\lambda+1)(2S'+1)(2S+1)}\, 
\nonumber \\ & & \times 
\threej{l'}{1}{l}{0}{0}{0} 
\sixj{\lambda'}{1}{\lambda}{l}{L}{l'} 
\sixj{S'}{1}{S}{\frac{1}{2}}{\frac{1}{2}}{\frac{1}{2}} 
\sixj{S'}{1}{S}{\lambda}{J}{\lambda'}\, 
\nonumber \\ & & \times 
\roundbraket{P'L}{NL} \roundbra{p'l'}v(r)\roundket{nl} \,, 
\end{eqnarray} 
\begin{eqnarray} 
\lefteqn{ 
\roundbra{p'l'\,P'L\,\lambda'S'J} i v(r) (\bold{\sigma}_1\times\bold{\sigma}_2)  
\cdot \hat{\bold{r}}  
\roundket{nl\,NL\,\lambda SJ} = 
} 
\nonumber \\ & & 
(-)^{L+J+S}\, \left(\delta_{S'0}\delta_{S1} + \delta_{S'1}\delta_{S0}\right)\, 
\sqrt{12 (2l'+1)(2l+1)(2\lambda'+1)(2\lambda+1)}\, 
\nonumber \\ & & \times 
\threej{l'}{1}{l}{0}{0}{0} 
\sixj{\lambda'}{1}{\lambda}{l}{L}{l'} 
\sixj{S'}{1}{S}{\lambda}{J}{\lambda'}\, 
\roundbraket{P'L}{NL} \roundbra{p'l'}v(r)\roundket{nl} \,, 
\end{eqnarray}
\begin{eqnarray} 
\lefteqn{ 
(-)^{S'+S}\, 
\roundbra{p'l'\,P'L\,\lambda'S'J} v(r)\bold{\sigma}_1 \cdot \bold{\nabla}  
\roundket{nl\,NL\,\lambda SJ} = 
} 
\nonumber \\ 
\lefteqn{ 
\rule{1.\thiswidth}{0cm} 
\roundbra{p'l'\,P'L\,\lambda'S'J} v(r)\bold{\sigma}_2 \cdot \bold{\nabla}  
\roundket{nl\,NL\,\lambda SJ} = 
}
\nonumber \\ & & 
(-)^{L+J+1}\, 
\sqrt{6 (2l'+1)(2l+1)(2\lambda'+1)(2\lambda+1)(2S'+1)(2S+1)}\, 
\nonumber \\ & & \times 
\threej{l'}{1}{l}{0}{0}{0} 
\sixj{\lambda'}{1}{\lambda}{l}{L}{l'} 
\sixj{S'}{1}{S}{\frac{1}{2}}{\frac{1}{2}}{\frac{1}{2}} 
\sixj{S'}{1}{S}{\lambda}{J}{\lambda'}\, 
\nonumber \\ & & \times 
\roundbraket{P'L}{NL} 
\roundbra{p'l'}v(r)\hat{d}_V(l',l;r)\roundket{nl} \,, 
\end{eqnarray} 
\begin{eqnarray} 
\lefteqn{ 
\roundbra{p'l'\,P'L\,\lambda'S'J} i v(r) (\bold{\sigma}_1\times\bold{\sigma}_2)  
\cdot \bold{\nabla}  
\roundket{nl\,NL\,\lambda SJ} = 
} 
\nonumber \\ & & 
(-)^{L+J+S}\, \left(\delta_{S'0}\delta_{S1} + \delta_{S'1}\delta_{S0}\right)\, 
\sqrt{12 (2l'+1)(2l+1)(2\lambda'+1)(2\lambda+1)}\, 
\nonumber \\ & & \times 
\threej{l'}{1}{l}{0}{0}{0} 
\sixj{\lambda'}{1}{\lambda}{l}{L}{l'} 
\sixj{S'}{1}{S}{\lambda}{J}{\lambda'}\, 
\nonumber \\ & & \times 
\roundbraket{P'L}{NL} 
\roundbra{p'l'}v(r)\hat{d}_V(l',l;r)\roundket{nl} \,, 
\end{eqnarray}
\begin{eqnarray} 
\lefteqn{ 
(-)^{S'+S}\, 
\roundbra{p'l'\,P'L\,\lambda'S'J} v(r)\bold{\sigma}_1 \cdot \bold{l}  
\roundket{nl\,NL\,\lambda SJ} = 
} 
\nonumber \\ 
\lefteqn{ 
\rule{1.\thiswidth}{0cm}
\roundbra{p'l'\,P'L\,\lambda'S'J} v(r)\bold{\sigma}_2 \cdot \bold{l}  
\roundket{nl\,NL\,\lambda SJ} = 
}
\nonumber \\ & & 
(-)^{l+L+J+1}\, \delta_{l'l} 
\sqrt{6l(l+1)(2l+1)(2\lambda'+1)(2\lambda+1)(2S'+1)(2S+1)}\, 
\nonumber \\ & & \times 
\sixj{\lambda'}{1}{\lambda}{l}{L}{l'} 
\sixj{S'}{1}{S}{\frac{1}{2}}{\frac{1}{2}}{\frac{1}{2}} 
\sixj{S'}{1}{S}{\lambda}{J}{\lambda'}\, 
\nonumber \\ & & \times 
\roundbraket{P'L}{NL} \roundbra{p'l'}v(r)\roundket{nl} \,, 
\end{eqnarray} 
\begin{eqnarray} 
\lefteqn{ 
\roundbra{p'l'\,P'L\,\lambda'S'J} i v(r) (\bold{\sigma}_1\times\bold{\sigma}_2)  
\cdot \bold{l}  
\roundket{nl\,NL\,\lambda SJ} = 
} 
\nonumber \\ & & 
(-)^{l+L+J+S}\, \delta_{l'l} 
\left(\delta_{S'0}\delta_{S1} + \delta_{S'1}\delta_{S0}\right)\, 
\sqrt{12 l(l+1)(2l+1) (2\lambda'+1)(2\lambda+1)}\, 
\nonumber \\ & & \times 
\sixj{\lambda'}{1}{\lambda}{l}{L}{l'} 
\sixj{S'}{1}{S}{\lambda}{J}{\lambda'}\, 
\roundbraket{P'L}{NL} \roundbra{p'l'}v(r)\roundket{nl} \,, 
\end{eqnarray}
\begin{eqnarray} 
\lefteqn{ 
\roundbra{p'l'\,P'L\,\lambda'S'J}v(r) \hat{\bold{r}}\cdot\bold{\nabla}
\roundket{nl\,NL\,\lambda SJ} = 
}
\nonumber \\ & & 
\roundbra{p'l'\,P'L\,\lambda'S'J}v(r)\hat{d}_S(r) 
\roundket{nl\,NL\,\lambda SJ} = 
\nonumber \\ & & 
\delta_{l'l} \delta_{\lambda'\lambda} \delta_{S'S} \roundbraket{P'L}{NL} 
\roundbra{p'l'}v(r)\hat{d}_S(r)\roundket{nl} \,, 
\end{eqnarray}
\begin{eqnarray} 
\lefteqn{ 
\roundbra{p'l'\,P'L\,\lambda'S'J} 
v(r)\, \bold{\sigma}_1\cdot\bold{\sigma}_2\, \hat{\bold{r}}\cdot\bold{\nabla}
\roundket{nl\,NL\,\lambda SJ} = 
}
\nonumber \\ & & 
\roundbra{p'l'\,P'L\,\lambda'S'J}
v(r)\, \bold{\sigma}_1\cdot\bold{\sigma}_2\,  \hat{d}_S(r) 
\roundket{nl\,NL\,\lambda SJ} \,,
\end{eqnarray}
\begin{eqnarray} 
\lefteqn{ 
\roundbra{p'l'\,P'L\,\lambda'S'J} 
v(r)\, \bold{\sigma}_1\cdot\hat{\bold{r}}\, \bold{\sigma}_2\cdot\bold{\nabla}
\roundket{nl\,NL\,\lambda SJ} = 
}
\nonumber \\ & & {} 
\frac{1}{3}
\roundbra{p'l'\,P'L\,\lambda'S'J}
v(r)\, \bold{\sigma}_1\cdot\bold{\sigma}_2\,  \hat{d}_S(r) 
\roundket{nl\,NL\,\lambda SJ} 
\nonumber \\ & & {} 
+ \frac{1}{3} 
\roundbra{p'l'\,P'L\,\lambda'S'J}
v(r)\, S_{12}(\hat{\bold{r}})\,  \hat{d}_T(l',l;r) 
\roundket{nl\,NL\,\lambda SJ}
\nonumber \\ & & {} 
+ \frac{1}{2} 
\roundbra{p'l'\,P'L\,\lambda'S'J}
i\frac{v(r)}{r}\, (\bold{\sigma}_1\times\bold{\sigma}_2)\cdot\bold{l}  
\roundket{nl\,NL\,\lambda SJ} \,,
\end{eqnarray}
\begin{eqnarray} 
\lefteqn{ 
\roundbra{p'l'\,P'L\,\lambda'S'J} 
v(r)\, \bold{\sigma}_2\cdot\hat{\bold{r}}\, \bold{\sigma}_1\cdot\bold{\nabla}
\roundket{nl\,NL\,\lambda SJ} = 
}
\nonumber \\ & & {} 
\frac{1}{3}
\roundbra{p'l'\,P'L\,\lambda'S'J}
v(r)\, \bold{\sigma}_1\cdot\bold{\sigma}_2\,  \hat{d}_S(r) 
\roundket{nl\,NL\,\lambda SJ} 
\nonumber \\ & & {} 
+ \frac{1}{3} 
\roundbra{p'l'\,P'L\,\lambda'S'J}
v(r)\, S_{12}(\hat{\bold{r}})\,  \hat{d}_T(l',l;r) 
\roundket{nl\,NL\,\lambda SJ}
\nonumber \\ & & {} 
- \frac{1}{2} 
\roundbra{p'l'\,P'L\,\lambda'S'J}
i\frac{v(r)}{r}\, (\bold{\sigma}_1\times\bold{\sigma}_2)\cdot\bold{l}  
\roundket{nl\,NL\,\lambda SJ} \,. 
\label{nmef} 
\end{eqnarray}

We have introduced the following effective radial differential operators:
\begin{eqnarray} 
\hat{d}_S(r) &=& \frac{\partial}{\partial r} \,, 
\nonumber \\
\hat{d}_V(l',l;r) &=& \frac{\partial}{\partial r} + 
\frac{l(l+1) - l'(l'+1) + 2}{2r} \,, 
\nonumber \\ 
\hat{d}_T(l',l;r) &=& \frac{\partial}{\partial r} + 
\frac{l(l+1) - l'(l'+1) + 6}{4r} \,. 
\label{diff} 
\end{eqnarray}
When the short range correlations are implemented as indicated in 
Eqs.~\Ref{src}, one should, accordingly, make the following 
replacements in the relative radial matrix elements in  
Eqs.~\Ref{nmei}--\Ref{nmef}:
\begin{eqnarray}
\roundket{nl} &\to& 
g_{\Lambda N}(r)\roundket{nl} \,, 
\nonumber \\
\roundbra{p'l'} &\to& 
\roundbra{p'l'}\, g_{NN}(r) \,. 
\end{eqnarray}
Clearly, the correlation function $g_{\Lambda N}(r)$ is also subject to the 
action of the differential operators in these equations.
Explicitly, the center-of-mass radial overlap and the relative radial matrix 
elements are given by 
\begin{equation}\label{overlap}
\roundbraket{P'L}{NL} = \int R^2\, dR\, j_L(P'R)\,  
\mathcal{R}_{NL}\!\left(b/\sqrt{2},\,R\right)  
\end{equation}
and
\begin{equation}\label{radial}
\roundbra{p'l'}v(r)\hat{d}(r)\roundket{nl} = 
\int r^2\, dr\, j_{l'}(p'r)\, g_{NN}(r)\, v(r)\, 
\hat{d}(r) \left[ g_{\Lambda N}(r)\, 
\mathcal{R}_{nl}\!\left(\sqrt{2}\,b,\,r\right) \right]  \,, 
\end{equation} 
where $j_L$ and $j_{l'}$ are spherical Bessel functions, 
$\mathcal{R}_{NL}$ and $\mathcal{R}_{nl}$  are harmonic oscillator radial 
wave-functions with the first arguments giving the respective length parameters, 
and $\hat{d}(r)$ is either unity or one of the differential operators 
\Ref{diff}.

Finally, let us remark that the convention for the relative coordinate adopted 
here, Eq.~\Ref{relative}, has the opposite sign to those of Refs. \cite{Ba02} 
and \cite{Mo59}, for instance.   
This introduces an extra phase factor of $(-1)^l$ and $(-1)^{l'}$, 
respectively, in our kets and bras involving the relative motion. 
As a result, the transformation brackets in Eq.~\Ref{moshinsky} 
differ by a factor of $(-1)^l$ from those originally defined in 
Ref.~\cite{Mo59}. 
To avoid this adjustment, one can simply shift to the opposite convention for 
the relative coordinate by making the transcriptions $\bold{r} \to - \bold{r}$, 
$\hat{\bold{r}} \to - \hat{\bold{r}}$ and $\nabla \to - \nabla$ in the 
expressions for the transition potential in coordinate space given in 
Subsections \ref{nstr} and \ref{str}. In either case, the equations in 
Subsection~\ref{rates} and here in Appendix~\ref{nuclear} remain formally 
unaltered. 

\section{Phase conventions for $a$, $b$, \dots $f$ \label{conventions}}
\appendixeqn

With the phase conventions of Nabetani \etal {} (N.O.S.K.), in Ref.~\cite{Na99}, 
from which Eq.~\Ref{asym} is obtained, the transition amplitudes are defined as 
follows: 
\begin{eqnarray}
a &=& \left(
\bra{pn, ^1\!\mathrm{S}_0}\hat{V}\ket{p\Lambda, ^1\!\mathrm{S}_0}
\right)_{\mathrm{N.O.S.K.}}  \,, 
\nonumber \\  
b &=& \left(
\bra{pn, ^3\!\mathrm{P}_0}\hat{V}\ket{p\Lambda, ^1\!\mathrm{S}_0}
\right)_{\mathrm{N.O.S.K.}}  \,, 
\nonumber \\ 
c &=& \left(
\bra{pn, ^3\!\mathrm{S}_1}\hat{V}\ket{p\Lambda, ^3\!\mathrm{S}_1}
\right)_{\mathrm{N.O.S.K.}} \,, 
\nonumber \\ 
d &=& \left(
\bra{pn, ^3\!\mathrm{D}_1}\hat{V}\ket{p\Lambda, ^3\!\mathrm{S}_1}
\right)_{\mathrm{N.O.S.K.}}  \,, 
\nonumber \\ 
e &=& \left(
\bra{pn, ^1\!\mathrm{P}_1}\hat{V}\ket{p\Lambda, ^3\!\mathrm{S}_1}
\right)_{\mathrm{N.O.S.K.}}  \,, 
\nonumber \\ 
f &=& \left(
\bra{pn, ^3\!\mathrm{P}_1}\hat{V}\ket{p\Lambda, ^3\!\mathrm{S}_1}
\right)_{\mathrm{N.O.S.K.}}  \,. 
\label{nosk}
\end{eqnarray}
The relationship with our phase conventions is 
\begin{equation}\label{phases}
\left( 
\bra{pn,\, ^{2S'+1}{l'}_J} \hat{V} \ket{p\Lambda,\, ^{2S+1}{l}_J}
\right)_{\mathrm{N.O.S.K.}} = (-)^{S'+S} \, i^{-l'} \, 
\bra{np,\, ^{2S'+1}{l'}_J} \hat{V} \ket{\Lambda p,\, ^{2S+1}{l}_J} \,,
\end{equation}
where the first correction is due to the change in ordering in the 
Clebsch-Gordan couplings for the spins, and the second one, to the fact that we 
do not include the phase $i^{l'}$  in the final partial-wave radial function 
(\emph{cf.} Eq.~\Ref{radial}). This explains the extra phase factors appearing 
in Eq.~\Ref{tramps}.

Notice that there is no factor $(-)^{l'+l}$ in Eq.~\Ref{phases} as might be 
expected due to the change in ordering of the two particles in both the initial 
and final states, the reason being that this is compensated by the fact that we 
use the opposite convention for the relative coordinates.

\end{appendix}

%

%
\end{document}